\pdfoutput=1
\documentclass{aastex6}

\begin{document}

\title{MAGNETIC FIELDS IN ACCRETION DISKS: A REVIEW}

\author{Amir Jafari}
\affil{Dept. of Physics \& Astronomy, Johns Hopkins University}

\begin{abstract}
We review the current theoretical models of the inward advection of the large scale external magnetic fields in accretion discs. The most plausible theories for launching astrophysical jets rely on strong magnetic fields at the inner parts of the host accretion disks. An internal dynamo can in principle generate small scale magnetic fields in situ but generating a large scale field in a disk seems a difficult task in the dynamo theories. In fact, as far as numerous numerical experiments indicate, a dynamo-generated field in general would not be coherent enough over the large length scales of order the disk's radius. Instead, a large scale poloidal field dragged in from the environment, and compressed by the accretion, provides a more promising possibility. The difficulty in the latter picture, however, arises from the reconnection of the radial field component across the mid-plane which annihilates the field faster than it is dragged inward by the accretion. We review the different mechanisms proposed to overcome these theoretical difficulties. In fact, it turns out, that a combination of different effects, including magnetic buoyancy and turbulent pumping, is responsible for the vertical transport of the field lines toward the surface of the disk. The radial component of the poloidal field vanishes at the mid-plane, which efficiently impedes reconnection, and grows exponentially toward the surface where it can become much larger than the vertical field component. This allows the poloidal field to be efficiently advected to small radii until the allowed bending angle drops to of order unity, and the field can drive a strong outflow.
\end{abstract}

\section{Introduction}

Outflows and jets, observed in some accretion disk systems, are most probably mediated by strong poloidal magnetic fields (Blandford \& Znajek 1977; Blandford \& Payne 1982; for a detailed review of jets see Pudritz et al. 2006). The jet launching mechanism is unlikely to depend on the nature of the central object since, while the jets are ubiquitous, the central accreting objects span a wide range of properties (Livio 1997; Ogilvie \& Livio 2001). The question then arises about the origin of these strong fields at the inner regions of the accretion disks. One attractive possibility is indeed internal magnetic dynamos. A dynamo, of course, can generate net magnetic flux only though interaction with the boundaries, the central object or outer boundary via flux expulsion, because of the flux conservation (Beckwith et al. 2009). Several dynamo models have been proposed based on the magneto-rotational and Parker instabilities and shear/rotation in the disks (see e.g., Tout \& Pringle 1996; Uzdensky \& Goodman 2008). Nevertheless, while disc dynamos can easily generate small scale magnetic fields, their capability is very restricted in generating large scale fields coherent over length scales of order the disk's radius. The proposed scenarios also remain speculative as far as the global disk simulations are concerned. On the other hand, numerical simulations imply that disks can drag the ambient fields, e.g., from the interstellar medium (ISM), inward toward the inner parts by the accreting matter (Suzuki \& Inutsuka 2014; Beckwith et al. 2009; Igumenshchev et al. 2003). In contrast to the numerical work, the magnetohydrodynamic (MHD) studies indicate that the field would diffuse away faster than it is dragged in (van Ballegooijen 1989; Lovelace, Romanova \& Newman 1994; Lubow, Papaloizou \& Pringle 1994; Lovelace, Newman \& Romanova 1997; Dyda et al 2013). Accretion discs are generally turbulent as a result of different instabilities most notably magneto-rotational instability (MRI). In such a turbulent medium, the assumption of flux freezing breaks down as a result of high turbulent diffusivity. The reconnection of the oppositely directed radial field components at the mid-plane annihilates the field at a rate larger than the advection rate. This was pointed out for the first time by van Ballegooijen (1989). Treating this as a simplified boundary condition problem, the author argued that the magnetic field in cataclysmic variable accretion disks would diffuse away over a time scale $\tau_{diff}=R/\alpha c_s$ much shorter than the accretion time scale $\tau_{acc}=R/|v_r|$ where $R$ is the disk's outer radius, $\alpha$ is Shakura-Sunyaev (1973) viscosity parameter, $c_s$ the speed of sound and $v_r$ the radial accretion velocity. 

One simple way to see the problem is to compare the inward advection rate with the outward diffusion rate. For the former, at a radius $r$, we can write $v_{adv}\sim \nu/r$ where $\nu$ is the turbulent viscosity. The bending of the magnetic field lines across the disk causes an outward diffusion. The corresponding azimuthal electric current is $J_\phi=v_{adv}B_z\sim \eta B_r/h$, with $h$ being the disk's height and $\eta$ the turbulent magnetic diffusivity, therefore $v_{diff}\sim (\eta/h)(B_r/B_z)$. In the stationary state, we require $v_{adv}\sim v_{diff}$ which, assuming that $\nu/\eta\sim 1$, leads to $B_r/B_z\sim h/r$ indicating a very small bending angle $B_r/B_z\ll 1$ (Guilet \& Ogilvie 2012).  

\begin{figure}
\includegraphics[scale=.25]{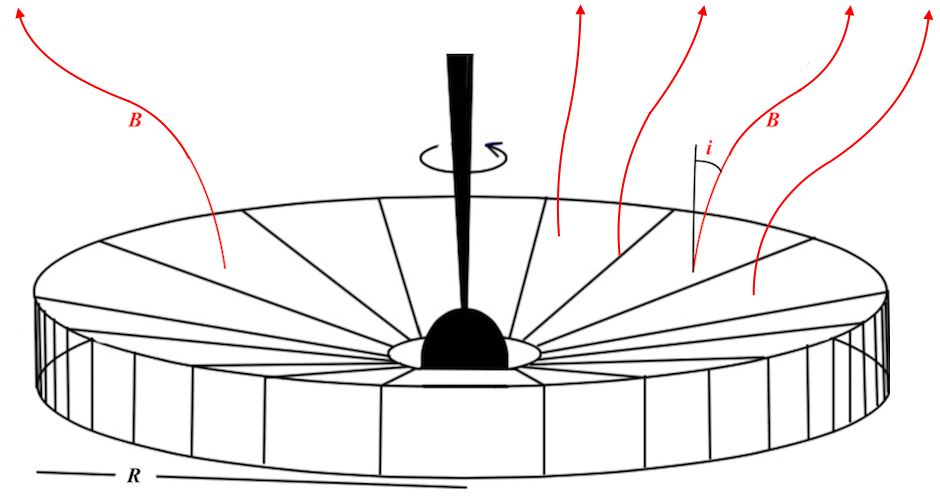}
\centering
\caption {\footnotesize {Fiducial magnetic field lines threading a thin accretion disc. The ambient magnetic field is usually assumed to be that of the interstellar medium ($B\sim 10^{-4}$G) possibly disturbed by a companion star. Launching outflows requires large bending angles $i=\tan^{-1}(B_r/B_z)\geq 30^\circ$. It is generally believed that these fields are dragged in from the outer boundary of the disk. (Illustration from Jafari \& Vishniac 2018)}}
\label{accretion-disc}
\end{figure}

This theoretical difficulty could be overcome, however, if the magnetic field were almost vertical in the disc since it would efficiently impede the reconnection of the large scale field and reduce the diffusion rate. Nevertheless, a vertical field cannot mediate outflows. In fact, a radial component of order the vertical field is necessary to launch efficient outflows and jets at the surface. Blandford and Payne (1982) proposed a necessary condition, to launch jets in accretion discs, in terms of the bending angle of the magnetic field lines as 
\begin{equation}\label{angle}
i=\tan^{-1}(B_r/B_z) \geq 30^\circ ,
\end{equation}
 where $B_r$ and $B_z$ are, respectively, the radial and vertical field components at the surface of the disc. Magnetic fields of at least of order 100 G seem necessary for this magneto-centrifugal mechanism to work (Blandford \& Payne 1982; for numerical simulations of jets see e.g., Romanova et al. 1997; Ouyed \& Pudritz 1999; Krasnopolsky et al. 1999). Such magnetic fields are usually assumed to be advected inward from the interstellar medium; see Fig.(\ref{accretion-disc}). It is worth to mention that  
the aforementioned magneto-centrifugal mechanism can operate even when the matter has a temperature less than the virial temperature or escape temperature (Ogilvie \& Livio 2001). Nevertheless, several authors have argued that some amount of thermal assistance would be still required to launch efficient outflows (e.g., Blandford \& Payne 1982; Ogilvie 1997; Ogilvie \& Livio 1998).

Bisnovatyi-Kogan \& Lovelace (2007) argued that the magnetic field lines are ``frozen in'' into the highly conducting surface layers of the disc where the strong magnetic fields and radiation flux suppress the turbulence; see \S \ref{SSBisnovatyi-Kogan2007}. In these radiative layers, the magnetic field diffuses much slower than the mid-plane where a fully developed turbulence is present because of the MRI (see also Bisnovatyi-Kogan \& Lovelace 2012). As a result, the field lines anchored to the surface tend to flow inward with almost the same speed as the accreting matter (see also Dyda et al. 2013). One serious issue with this picture is that it is very unlikely for the turbulence to be completely shutdown at the surface. Aside form the MRI, different hydrodynamic and plasma instabilities, e.g., Parker instability, would make these layers turbulent at least weakly. On the other hand, it is not clear how matter and the field could accrete inward without any turbulent viscosity which originates from turbulence. Lovelace et al. (2009) has have found stationary solutions to the MHD equations that implies that a large scale magnetic field does not diffuse away in an accretion disc with large plasma beta, $\beta>1$. In this rather peculiar picture, the flow is found to be radially outward near the mid-plane and radially inward in the parts near the surface. This study found that Prandtl numbers larger than a critical value $\sim 2.7$ can trigger magneto-centrifugal outflows from the surface. For smaller magnetic Prandtl numbers, this work has found electromagnetic outflows instead of jets. However, numerical simulations, in general, do not imply any sign of meridional circulation in discs. Only few other numerical works have indicated a similar picture (Zhu \& Stone 2013). Beckwith et al. (2009) found an almost similar patter in which there is an efficient inward field advection high in the corona while an outward motion drags the field out near the disc; see \S \ref{SSBeckwith2009}.

Spruit \& Uzdensky (2005) suggested a mechanism for the inward accretion of magnetic fields; see \S \ref{SSSpruit2005}. The authors suggest that the turbulent diffusion can be effectively reduced by grouping large scale vertical magnetic fields into small bundles through a mechanism known as magnetic flux expulsion. In these bundles, the fields are assumed to be strong enough to quench the underlying turbulence in order to avoid an outward diffusion (Guan \& Gammie 2009). This flux expulsion also occurs on the solar surface. The result of this mechanism is two-folded: first, it reduces the rate of magnetic reconnection, and second, it makes the loss of angular momentum much more efficient for the patches. The latter is the reason that the patches can accrete inward at a higher rate than that of the outward magnetic diffusion. The concentration of the magnetic flux around the central object leads to a bundle of the field lines. In such a disc, called magnetically arrested disc, this strong magnetized bundle will affect the accretion process (Narayan et al 2003; Spruit \& Uzdensky 2005). One question, however, arises regarding the stability of these magnetic bundles. These bundles require a long time to arrive at the central parts during which hydrodynamic and magnetohydrodynamic instabilities probably would destroy the supposed geometry.

Jafari \& Vishniac (2018) suggested a combination of different mechanisms to reduce the outward diffusion of the poloidal field; see \S \ref{SSJafari2017}. The key ingredient in their model is magnetic buoyancy that can prevent turbulent mixing of the field lines in the disk by transporting the field lines vertically toward the surface. The authors showed that turbulent density pumping, which competes with the buoyancy in the vertical field transport, has a relatively smaller effect in most realistic situations. The transport of the large scale magnetic field, in inhomogeneously turbulent fluids, by means other than the mean flow of matter is called pumping. The other similar mechanism involved in the vertical transport of the large scale field is turbulent diamagnetism. The latter effect too helps magnetic buoyancy in the vertical field transport. This effect originates from a gradient in turbulent magnetic diffusivity (Zeldovich 1957; Spitzer 1957). The partial suppression of turbulence at the surface layers of an accretion disc by the radiative flux (Bisnovatyi-Kogan \& Lovelace 2007) and a turbulent diffusivity at the mid-plane creates a diffusivity gradient that leads to a vertical drift of field lines. Both buoyancy and turbulent diamagnetism tend to hinder the reconnection of the radial field on the mid-plane which leads to an inefficient outward diffusion. 

\section{Disk Dynamics}

In an accretion disc with the density profile $\rho(r,z)$, the surface density, defined as $\Sigma(r)=\int_{-\infty}^{+\infty} \rho dz$ satisfies 
\begin{equation}\label{accretion disc-1}
r\frac{\partial \Sigma}{\partial t}+\frac{\partial}{\partial r}(r\Sigma v_r)=0.
\end{equation}
The conservation of angular momentum is governed by

\begin{equation}\label{accretion disc-4}
\frac{\partial }{\partial t}(r^2 \Omega \Sigma)+\frac{1}{r}\frac{\partial}{\partial r}(r^3 \Omega \Sigma v_r)=\frac{1}{r}\frac{\partial }{\partial r} \left( r^3\nu\Sigma \frac{ \partial \Omega }{ \partial r }  \right).
\end{equation}
In the steady state, the integration of the above expression leads to 
\begin{equation}
v_r={3\over 2}{\nu\over r}.
\end{equation} 
Even with a large turbulent viscosity, the radial velocity is much smaller than the Keplerian orbital velocity $v_r\lesssim10^{-3} v_K$. One can use equations (\ref{accretion disc-1}) and (\ref{accretion disc-4}) to write

\begin{equation}\label{accretion disc-5}
\frac{\partial \Sigma}{\partial t}=\frac{-1}{r}\frac{\partial}{\partial r} \left(    \frac{ 1 }{ \frac{\partial}{\partial r}(r^2\Omega)  }   \frac{\partial}{\partial r}(r^3\nu\Sigma \frac{\partial \Omega}{\partial r})    \right).
\end{equation}
This is the evolution equation for a general viscous accretion disc. For a Keplerian disc, the rotation profile is $\Omega=(GM_*/r^3)^{1/2}$ where $G$ is the gravitational constant and $M_*$ is the mass of the central star. For this particular rotation profile, the evolution equation reads
\begin{equation}\label{accretion disc-6}
\frac{\partial \Sigma}{\partial t}=\frac{3}{r}\frac{\partial}{\partial r} \left( \sqrt{r} \frac{\partial}{\partial r}(\nu \Sigma \sqrt{r})   \right),
\end{equation}
Combining eq.(\ref{accretion disc-1}) with eq.(\ref{accretion disc-6}), we can find the radial velocity as
\begin{equation}
  v_r={3\over \Sigma r^{1/2}}{\partial\over\partial r}(\nu\Sigma r^{1/2}).
  \end{equation}
Using $\partial M/\partial t=2\pi r \Sigma v_r$ yields
 \begin{equation}
{\partial M\over\partial t}=6\pi r^{1/2}{\partial\over\partial r}(\nu\Sigma r^{1/2}).
 \end{equation}
 
The accretion time scale is then given by
\begin{equation}\label{disk-evolution-time}
\tau_{acc}=\frac{R^2}{\nu},
\end{equation}
where $R$ is the approximate radius of the disk.

In the steady state, the rate of mass flow $\dot M$ is constant. Setting $\dot M=-2\pi r \Sigma v_r$ as a constant in eq.(\ref{accretion disc-4}) and assuming there is no torque we integrate from the inner boundary $r_0$ to $r$ to find
\begin{equation}\label{accretion disc-7}
\Sigma(r)=\frac{\dot M }{3\pi \nu }\left( 1-\sqrt{  \frac{r_0}{r}    }\right).
\end{equation}
This expression can be used to find the density profile as $\rho \propto  r^{-15/8}$. Eq.(\ref{accretion disc-7}) shows that far from the inner part of the disc, the mass flow rate is 
\begin{equation}\label{accretion disc-8}
\dot M|_{r=\infty}=3\pi \nu \Sigma |_{ r=\infty}.
\end{equation}
Hence the viscosity has a major role in the evolution of the disc and its mass distribution. The viscous time $\tau_{acc} \sim R^2/\nu$ is of order $10^{13}$ for a typical accretion disk at the radius $\sim 1\; AU$. This implies that there should be other type of viscosity which can more efficiently enhance the transport of angular momentum outward.

\subsection{Magnetorotational Instability (MRI)}
An accretion disk is a stable system from a hydrodynamical point of view if the Rayleigh's criterion is held:
\begin{equation}
{d(r^2\Omega)\over dr}>0,
\end{equation}
where $\Omega$ is the angular frequency. However, there are magnetohydrodynamic and plasma instabilities that disks are generally prone to. In the presence of a weak magnetic field, for instance, the stability condition is an $\Omega$ increasing with radius (Chandrasekhar 1961). This condition, however, is usually violated in accretion disks and leads to magnetorotational instability or MRI (Sano et al. 2004). MRI is the primary source of turbulence, which enhances angular momentum transport, in accretion disks (Balbus \& Hawley 1991). In order to get an insight, consider a parcel of fluid at some radius $r$ at which a coordinate system can be constructed with its $x$ axis directed radially outward and $y$ axis in the increasing azimuthal direction. The equations of motion read

\begin{equation}\label{MRI-1}
\ddot x-2\Omega_0 \dot y=-xr{d\Omega_0\over dr}-\Big({k^2B^2\over \rho} \Big)x,
\end{equation}
and
\begin{equation}\label{MRI-2}
\ddot y+2\Omega_0 \dot x=-\Big({k^2B^2\over \rho} \Big)y.
\end{equation}
The second term in the LHS of both equations represents the Coriolis force. The first term in the RHS of eq.(\ref{MRI-1}) is the result of the difference in the centripetal forces that the parcel experiences if slightly perturbed, by the radial displacement $x$, from its original orbit of radius $r$ with angular velocity $\Omega_0$; $r[\Omega_0^2(r)-\Omega_0^2(r+x)]$. The last terms in the RHS in both equations is a restoring magnetic force coming from the magnetic tension $(\bf{B.\nabla})\bf{B}/\rho$ where $\rho$ is the density. With a vertical magnetic field ${\bf{B}}=\hat k B$, the induction equation, $\delta {\bf{B}}\simeq\nabla\times({\bf{x\times B}})$ gives $\delta{\bf{B}}\sim ikB \bf{x}$ for a perturbation ${\bf{x}}={\bf{v}}\delta t$. The magnetic tension thus becomes $({\bf{B.\nabla}})\delta{\bf{B}}/\rho\sim (-k^2B^2/\rho)\bf{x}$. In order to get a dispersion equation we write $x,y \sim e^{i\omega t}$,
\begin{equation}
\omega^4-\Big( {2k^2B^2\over \rho}+\Big({k^2B^2\over\rho}\Big)^2 \Big)\omega^2+{k^2B^2\over\rho}\Big({k^2B^2\over\rho}+r{d\Omega^2\over dr}  \Big)=0.
\end{equation}
With a pure imaginary frequency $\omega$, the solutions $x,y \sim e^{i\omega t}$ will grow exponentially and the disc will become unstable. The necessary condition for this is $d\Omega^2/dr<0$ for the wavenumbers satisfying
\begin{equation}
{k^2B^2\over \rho}+r {d\Omega^2\over dr}<0.
\end{equation}
Obviously, a Keplerian disc is unstable for the MRI with a growth rate $\gamma=3\Omega/4$ (which is very rapid corresponding to an amplification factor of more than 100 per rotation period). The wavenumber satisfies the relation $k^2B^2/\rho=15\Omega^2/16$. Thus in a Keplerian disc threaded by a large scale vertical field $B_z$, the most unstable mode has the wavelength $\lambda_{mri}=2\pi \sqrt{(16/15)}V_{Az}/\Omega$ where $V_{Az}$ is the Alfv\'en velocity corresponding to the vertical field.

MRI is the most important mechanism responsible for turbulence in accretion discs. Its requirements are too easy to be satisfied so we expect discs are generally MRI-turbulent (for the numerical simulations see e.g., Brandenburg et al. 1995; Hawley et al. 1995b; Hawley 2001; Fromang \& Nelson 2006; Bai \& Stone 2013). Enhanced turbulent viscosity $\nu$ resolves the problem of extremely slow accretion rate with molecular viscosity. An enhanced magnetic diffusivity $\eta$ seems a necessary, however somehow oversimplified, ingredient in the study of turbulent conducting fluids. The magnetic Prandtl number is defined as the ratio of turbulent viscosity over turbulent magnetic diffusivity;
\begin{equation}
Pr_m={\nu\over \eta}.
\end{equation}

Numerical simulations of MRI in thin accretion discs imply a magnetic Prandtl number of order unity, $Pr_m\sim 1$ (see e.g., Guan \& Gammie 2009; Lesur \& Longaretti 2009; Fromang \& Stone 2009). This is while the microscopic (non-turbulent) magnetic Prandtl number can deviate from unity and have a different behavior (Balbus \& Hawley 2008). 

Majority of vertically stratified shearing-box simulations consider configurations with zero net vertical magnetic flux (Miller \& Stone 2000; Ziegler \& R{\"u}diger 2000; Hirose et al. 2006; Davis et al. 2010; Shi et al. 2010; Flaig et al. 2010; Simon et al. 2012). But in fact a net vertical magnetic field seems as an essential ingredient of the accretion discs which is explored rarely. It is believed that a turbulent stresses increase with increasing the net magnetic flux (Hawley et al. 1995b). In the recent years, more realistic shearing box simulations have been performed including a net flux with efficient outflows (Suzuki \& Inutsuka 2009 and Suzuki et al. 2010; Bai \& Stone 2013). Unstratified shearing box simulations imply that the turbulent stress increases linearly with the inverse plasma $\beta$ at the mid-plane (Hawley et al. 1995b, Bai \& Stone 2011). This is expected to hold also in stratified shearing box simulations (Bai \& Stone 2013).

\subsection{Shakura-Sunyaev Model}
The so-called "turbulent magnetic diffusivity" is an attempt to consider the effect of turbulence on magnetic diffusivity. Since electrical resistivity originates from particle collisions, one may think of an "enhanced" resistivity as a result of turbulent motions. In fact, a mere replacement of magnetic diffusivity with turbulent diffusivity is indeed illegitimate but a more complicated consideration based on the concept of magnetic helicity is helpful in formulating such a quantity. This said, we can perform a simple dimensional analysis employing the turbulent eddy length scale $l$ and eddy velocity $\delta v$ in the turbulent cascade, to get a rough estimate of the turbulent diffusivity as $\eta_t\sim l \delta v$. In fact, the turbulent diffusivity is known as the $\beta$-effect in the mean field dynamo theories and is given by 
\begin{equation}
\eta_t\equiv\beta\simeq {\delta v^2\over \tau}.
\end{equation}
with eddy turn-over time $\tau$. There is a similar problem, in turbulent media, with the kinematic viscosity which is too small in accretion disks to provide the observed accretion rates. The turbulent viscosity given by eq.(\ref{Shakura-Sunyaev-1}) is, however, orders of magnitude larger than kinematic viscosity. In this model, viscous stress tensor $T_{r\phi}$ is assumed to be proportional to the pressure $P=\rho c_s^2$ with $\rho$ being the gas pressure and $c_s$ the speed of sound. Since the azimuthal velocity $v_\phi$ is much larger than the radial, $v_r$, and vertical velocity, $v_z$, we can write
\begin{equation}
T_{r\phi}=-\rho\nu\Big({1\over r}{\partial v_r\over\partial\phi}+{\partial v_\phi\over\partial r}-{v_\phi\over r}\Big)=-\rho\nu r{\partial \Omega\over \partial r}.
\end{equation}
Thus, the assumption $T_{r\phi}=\alpha P$ leads to the following well-known ansatz:
\begin{equation}\label{Shakura-Sunyaev-1}
\nu=\alpha_{ss}c_sh=\alpha_{ss} {c_s^2\over \Omega}=\alpha_{ss} {P_{tot}\over \rho\Omega},
\end{equation}
where $h$ is the height of the disc and $\alpha \leq 1$ is introduced as a dimensionless parameter. A typical value for $\alpha$ in proto-planetary disks is $\sim 0.01$, while for the accretion disks around compact objects $\alpha\sim 0.1$. Since the sound speed in the disc roughly equals $h\Omega$, so we find $\nu(r)\simeq \alpha \Omega h^2$ which increases with radius as $r^{1/2}$ provided that $h/r\ll 1$ is assumed to be constant. One important advantage of $\alpha$-prescription is to confine all our uncertainties about the stresses into a single parameter called $\alpha$ (Sano et al. 2004). In ideal MHD simulations, the saturation value of $\alpha$ varies between $0.001$ and $0.1$ (Hawley et al. 1995;1996). 

In fact, $\alpha$ is unlikely to be a constant, rather it depends on the strength and geometry of the magnetic field (Pessah et al. 2007). A major difficulty is to distinguish the numerical and physical dependences (see e.g., Hawley et al 1996). Despite this, order of magnitude calculations and numerical simulations indicate that $\alpha$ increases with increasing the magnetic field. For example, Sano et al. (2004) found that $\alpha$ was proportional to $B_z^{3/2}$. Pessah et al. (2007) and Bai and Stone (2013) found that $\alpha$ increased monotonically with the net vertical field. Zhu and Stone (2017) found a dependence on radius as $\alpha\sim R^{-2/5}$ for the vertically integrated alpha which is larger than its mid-plane value by a factor of $10$.  In the latter work, dynamo action is suppressed in the presence of strong vertical fields ($\beta<1000$ at the mid-plane). However, neither the dependence on radius nor on the magnetic field is yet well-established.

\section{Magnetic Fields and Outflows}
The most plausible mechanism for launching astrophysical jets, emerging from some accretion discs, involves strong magnetic fields somehow get concentrated in the inner regions near the accreting mass (Blandford \& Znajek 1977; Blandford \& Payne 1982). The origin of these magnetic fields, however, remains unclear. One simple solution is to assume an internal dynamo generating the required magnetic field in the disc. The major difficulty with this proposal is the strength, length scale and configuration of the dynamo-generated fields which are generally inappropriate for launching jets (see e.g., Burm \& Kuperus 1988; Tout \& Pringle 1996). Another possibility is trapping the ambient interstellar magnetic field and advecting it while it is frozen into the accreting matter. Numerical simulations seem to confirm such an inward transport of the magnetic fields threading the disc, by the accreting matter, from the environment (Suzuki \& Inutsuka 2014; Beckwith et al. 2009; Igumenshchev et al. 2003). Yet the involved mechanism of this transport is not fully understood since there are theoretical difficulties that make the accretion of magnetic field along with the matter an implausible picture. Accretion discs are generally turbulent as a result of magneto-rotational instability (MRI) which breaks the notion of flux freezing. Thus the magnetic field cannot get tightly frozen into the accreting plasma and, because of the enhanced diffusivity, there would be an outward diffusion of the magnetic field. The other difficulty stems from the reconnection of the oppositely directed radial field components at the mid-plane. These issues, may be resolved if the magnetic field were almost vertical in the disc which would efficiently reduce the reconnection rate making the diffusion time scale independent of the disc's height. Nevertheless, this leads to another difficulty: in order to have efficient outflows, the magnetic field at the surface must make a $60^\circ$ angle with the vertical (Blandford and Payne 1982),

\begin{equation}\label{angle}
i=\tan^{-1}\Big({B_r\over B_z}\Big) \geq 30^\circ ,
\end{equation}
 where $B_r$ and $B_z$ are, respectively, the cylindrical radial and vertical field components at the surface of the disc. To see how this condition arises let us consider a Keplerian disc with angular velocity $\Omega=(GM_*/r^3)^{1/2}$ with $M_*$ being the central accreting mass. Assuming flux freezing, any magnetic field line will rotate with the disc with an angular velocity $\Omega_0=(GM_*/r_0^3)^{1/2}$ where $r_0$ is the radius the field line touches the mid-plane. A parcel of plasma at $(r_0, z)$ loaded onto and co-rotating with a bow-shaped field line whose foot passes through the mid-plane at $(r_0,0)$ has an effective potential (per mass) given by
\begin{equation}\label{potential}
\Phi(r,z)=-{GM_*\over r_0}  \Big({r_0\over \sqrt{r^2+z^2}}+{1\over 2} ( {r\over r_0} )^2  \Big),
\end{equation}
The Taylor expansion of this potential, to the second order, is given by
\begin{equation}
\Phi(r+\Delta r,\Delta z)\simeq\Phi(r_0, z=0)-{GM_*\over r_0}  \Big( {3\Delta r^2\over 2r_0^2}   - {\Delta z^2\over 2r_0^2}\Big),
\end{equation}
which, requiring $\Phi(r+\Delta r,\Delta z)\simeq\Phi(r_0, z=0)$ gives the stability condition as 
\begin{equation}
\Big({\Delta z\over \Delta r}\Big)^2=3 \rightarrow \tan \Big({B_z\over B_r}\Big)=\pm {\pi\over 3}.
\end{equation}

\begin{figure}
\includegraphics[scale=.35]{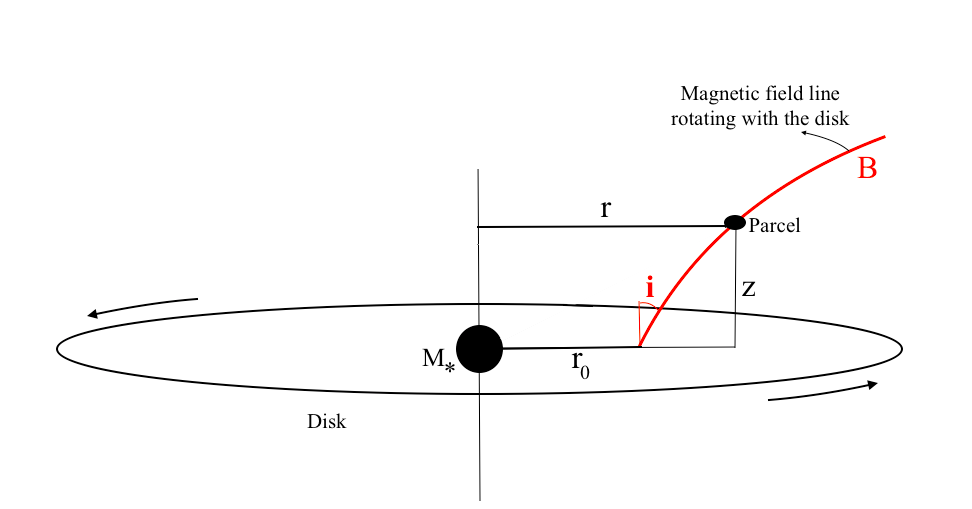}
\centering
\caption {\footnotesize {Magneto-centrifugal mechanism proposed by Blandford \& Payne (1982): A parcel of plasma, at radius $r$ and height $z$ above the disk, moving along a magnetic field line whose footpoint on the disk lies at the radius $r_0$. The field line and the parcel rotate with rotational velocity $\Omega_0=(GM_*/r_0^3)^{1/2}$ so the parcel's velocity is $v=r\Omega_0$ . Matter can move along the magnetic field lines co-rotating with the disk in a similar manner that a bead would move on a string when it experiences a centrifugal force along the string. The effective potential for the parcel is given by eq.(\ref{potential}). This potential is basically the sum of the gravitational potential energy and the "effective kinetic energy" per mass of the parcel, the latter given by $v^2/2=(r\Omega_0)^2/2=(GM_*/2r_0)(r/r_0)^2$. The parcel can accelerate along the field lines provided that the field has a large bending angle; $i\geq 30^\circ$.}}
\label{magneto}
\end{figure}

This result shows that the angle between the isopotential line and the disc's rotational axis at $(r_0,0)$ is $i=\pm \pi/6$. Therefore, matter can move along any force line inclined by this angle either inward or outward. If $i<30^\circ$, the matter can go up along the effective potential but extra energy is required to launch it from the surface. If $i>30^\circ$, the matter can go down on the effective potential without requiring more energy to be expelled from the disc. Although the mass of the accreting black hole, or star, may determine the jet's velocity (King et al. 2015), but magnetic fields are critical in collimation and launching these energetic particles. Magnetic fields of at least of order 100 G seem necessary for this magneto-centrifugal mechanism to work (Blandford \& Payne 1982). Such magnetic fields are usually assumed to be advected inward from the interstellar medium; see Fig.(\ref{magneto}). 

In a highly conductive accretion disc, the force lines would be tightly anchored to the matter. As a result, the inward flow of the matter could transport the poloidal magnetic field toward the central mass as well. Accretion discs actually seem to drag poloidal magnetic fields inward, or at very least, prevent it from diffusing outward if it gets concentrated at the inner regions (Beckwith, Hawley \& Krolik 2009). The enhanced turbulent diffusivity breaks the flux freezing condition and the field diffuses away relatively fast (Park \& Vishniac 1996). Magnetohydrodynamic studies of the turbulent discs indicate that a weak large scale magnetic field threading an accretion disc diffuses away rapidly (van Ballegooijen 1989; Lovelace, Romanova \& Newman 1994; Lubow, Papaloizou \& Pringle 1994a; Lovelace, Newman \& Romanova 1997; Dyda et al 2013). The outward diffusion of the poloidal field would be faster than the inward accretion velocity for a magnetic Prandtl number of order unity $P_m \sim 1$ which is a widely accepted value for the thin discs.

In principle, there are two ways to provide a strong magnetic field near the accreting mass. One is an internal dynamo in the disc and the other is transporting the interstellar field from the outer edge to the inner parts. Both these mechanisms have their own theoretical difficulties. Although an internal dynamo may seem as a potential mechanism to compensate the outward diffusion of the magnetic field in a disc by generating a large scale field with no help from the ambient field (Guan \& Gammie 2009). Nevertheless, the strength and configuration of such a dynamo-generated-field seem inappropriate for launching jets; see below. The second mechanism, transporting the external field with the accreting matter, encounters another difficulty first pointed out by van Ballegooijen (1989). The azimuthal electric current of order $J_\phi\sim B_r/H$ combined with the Ohm's law, $v_{diff}\times B_z\sim \eta J_\phi$ indicates a diffusion rate of order $v_{diff}\sim (\eta/H)(B_r/B_z)$. If the outward diffusion of the field has to be balanced by its inward advection then $\nu/R\sim (\eta/H)(B_r/B_z)$. Efficient outflows at the surface requires $B_r/B_z\sim 1$ near the surface so $\nu/\eta\sim R/H$ in contrast with the generally accepted  assumption that $Pr_m=\nu/\eta=1$. There are several remedies to this theoretical difficulty which we will study in the next subsections.

\subsection{Dynamo Action}

The simplest way, to explain the strong fields, presumed to be present near the discs' inner parts, is to consider in situ generation of the magnetic field by an internal dynamo in the disc. Nevertheless, it is important to emphasis that magnetic field is divergence-free and the net flux is conserved in the disc so any dynamo could only work through interactions with the boundaries (Beckwith et al. 2009). A simple dynamo action can be thought of as trapping the supposedly vertical field $B_z$ threading the disc from the interstellar medium by conductive accreting matter. This component is dragged inward by the accreting matter and stretched from the vertical generating a radial component $B_r$. Differential rotation and shear will then generate a toroidal component $B_\phi$. To complete the cycle, there has to be a mechanism producing a vertical field out the latter two components. Tout and Pringle (1992) suggested that, apart from the shearing the radial field, the Parker instability (magnetic buoyancy), reconnection and Balbus-Hawley instability (MRI) could in principle produce a self-sustaining dynamo action in an isothermal disc. Their simple model does not depend on any initial turbulence to begin with. The generation of the toroidal field out of the radial field is in fact an $\omega$-dynamo process which is also a part of the MRI producing $B_\phi$ and $B_r$ simultaneously from a vertical field. The shear energy is converted into the magnetic energy of the azimuthal field which itself is unstable to the buoyancy. With a fixed angular velocity profile $\Omega(r)$, following Tout and Pringle (1992), we write
\begin{equation}\label{tout-pringle1992-1}
{dB_\phi\over dt}={3\over 2} \Omega B_r-{B_\phi\over \tau_P}.
\end{equation}
The first term in the RHS indicates the gaining a toroidal component from a radial field while the second term indicates the loss loss of the toroidal component by the Parker instability, with the growth time scale $\tau_P$. In the equilibrium state of the dynamo, $B_z$ is generated by $B_\phi$ while annihilated by reconnection with a time scale $\tau_R$:
\begin{equation}\label{tout-pringle1992-2}
{dB_z\over dt}={B_\phi\over \tau_P}-{B_z\over \tau_R}.
\end{equation}
The radial component is a little complicated but here, as well, the loss is attributed to the Parker instability. However, the radial component also gets sheared which imposes a real difficulty. Some authors have argued that the shear can enhance (see Coroniti 1981) or diminish the Parker instability (Vishniac \& Diamond 1992). One also needs to take into account the effect of the MRI in order to write the time evolution for the radial component. 

\begin{equation}
{dB_r\over dt}=\gamma \Omega B_z-{B_r\over\tau_P}.
\end{equation}
Here, $\gamma$ is a constant which depends on the maximum growth rate of the MRI and the magnitude of the vertical magnetic field (see Tout \& Pringle 1992 for details). With estimated time scales for the Parker instability and magnetic reconnection, one should look for the equilibrium when $dB_z/dt=dB_\phi/dt=dB_r/dt=0$. Doing so, Tout and Pringle (1992) showed the dynamo action was unstable to the presence of a small scale seed field which get amplified over the shear time scale. Since the MRI has a maximum allowed magnitude for $B_z$ in order to operate in the disc, so this dynamo growth would finally saturate. In fact, shear is the ultimate mechanism that derives the dynamo and hence gives rise to an enhanced magnetic diffusivity. Tout and Pringle (1992) predicted a dynamo generated field $B_d$ of coherence scale of $H$, the disc's half-thickness, that in a gas pressure dominated disc becomes of order
\begin{equation}
B_d\sim \sqrt{4\pi \rho} c_s,
\end{equation}
where $c_s$ is the speed of sound and $\rho$ is the average density. These latter results are also confirmed in several other numerical simulations (Brandenburg et al. 1995; Hawley et al. 1995a; and Matsumoto \& Tajima 1995). The predicted plasma-$\beta$ is orders of magnitude smaller than the generally accepted value which is of order $\beta\sim 10^2$ (cf. Bai \& Stone 2013). More importantly, magnetic fields required for jets should have a length scale of order the disc's radius $R$ which is much larger than $H$. In fact, an order of magnitude calculation (Pringle 1993) shows that a minimum requirement for a vertical field threading the disc is 
\begin{equation}
{B_z^2\over B_d^2}\sim {\dot M_{jet}\over \dot M_{acc}}{H\over R},
\end{equation}
where $\dot M_{jet}$ is the mass loss through the jet and $\dot M_{acc}$ is the accretion rate. One solution to obtain such large length scales for the magnetic field launching the jet from the dynamo-generated fields with much smaller length scale, is magnetic reconnection (Fricsh et al. 1975) or an inverse cascade (Tout \& Pringle 1996). 

Several numerical simulations have been performed starting with a zero initial poloidal field in the disc (De Villiers et al. 2003) where some initial dipole loops are contained within an isolated plasma torus. In these simulations as well, the differential rotation rapidly generates a toroidal component which is strong enough in the plunging region of the flow to eject a magnetic tower into the funnel region. This in turn leads to a large scale dipole field which anchors the central black hole. In the case of a Kerr black hole, this dipole field can produce a Poynting flux jet. Other than the initially dipolar fields, other alternative configurations such as quadrupole field loops have also been considered (e.g., McKinney \& Gammie 2004; McKinney \& Blandford 2009). Beckwith et al. (2008) studied non-dipolar initial field loops and also models initially beginning with purely toroidal fields. In the latter case, the toroidal field generates the MRI that in turn creates turbulence (Beckwith et al. 2009). One common feature of many of these dynamo models (the so-called $\alpha\omega$ dynamos) is an "upward drift" of the magnetic flux by some mechanism such as turbulence or convection (Camenzind 1994; Stepinski 1995) or magnetic buoyancy (Tout \& Pringle 1992). None of the proposed sophisticated dynamo models (see e.g., Camenzind 1994; Stepinski 1995; Uzdensky \& Goodman 2008; Beckwith et al. 2009) has been generally accepted so far as a mechanism to generate the magnetic fields required in launching powerful outflows. Astrophysical jets need strong large scale poloidal magnetic fields in the vicinity of the accreting mass (Blandford 1993). This is while most dynamo-generated fields have much smaller length scales often in the form of closed loops (Burm \& Kuperus 1988; Tout \& Pringle 1996).

Lubow and his collaborators (1994a) studied a case in which the total field is generated through both internal and external currents (that is a disc with a dynamo and threaded by an external field). They found a stationary solution for $B_z$ provided that ${\cal{D}}=(R/H)(\eta/\nu)\leq 1$. Since this condition is in contrast with the result of numerous numerical simulations showing that $\eta/\nu\sim 1$ so the authors argued that, other than turbulent viscosity, the winds could also contribute to the loss of angular momentum. So if the field is sufficiently strong to give considerable outflows, the angular momentum can be lost more efficiently leading also to a stationary solution for the magnetic field. Nevertheless, it is not clear how an efficient outflow can start initially with no strong magnetic field around and how it is to proceed in a self-sustaining manner (see e.g., Lubow et al. 1994b). 

\begin{figure}
\includegraphics[scale=.35]{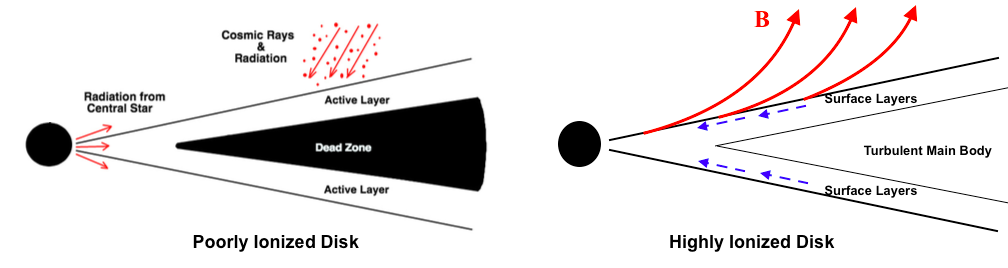}
\centering
\caption {\footnotesize {Non-turbulent surface layers hypothesized by Bisnovatyi-Kogan and Lovelace (2007) essentially leads to a "layered disk". Left: Cosmic rays and radiation can penetrate to some depth ionizing the matter and enforcing the MRI to operate in a poorly ionized disk such as a propto-planetary disk. The deeper layers across the mid-plane remain poorly ionized and so only weakly turbulent---a dead zone (Gammie 1996). Right: In a highly ionized and conductive disk, Bisnovatyi-Kogan and Lovelace suggest, the main body of the disk would be MRI-turbulent but radiation and magnetic fields would suppress the turbulence at the "surface layers". The authors have argued that this would lead to a complete shut-down of the turbulence in the surface layers allowing the magnetic field to be advected inward efficiently along with the accreting matter. However, the complete shut-down of the turbulence in a disk susceptible to different instabilities seems unlikely. Parker instability, among others, can operate near the surface generating turbulence. A closely related concept, magnetic buoyancy, seems to play a more important role in reducing the reconnection rate across the mid-plane; see \S \ref{SSJafari2017}. }}
\label{dead-zone}
\end{figure}

\subsection{Non-turbulent Surface Layers}\label{SSBisnovatyi-Kogan2007}

Bisnovatyi-Kogan \& Lovelace (2007; 2012) suggested that the magnetic field is ``frozen in'' into the highly conducting, and supposedly non-turbulent, surface layers of the disc. The argument goes as follows: at the surface layers of the disc, strong magnetic fields and radiation flux suppress the turbulence. This is similar to the suppression of the convection over the photospheres of stars with outer convective zones. In these radiative layers, the magnetic field diffuses much slower than where a fully developed turbulence is present (Bisnovatyi-Kogan \& Lovelace 2007). Consequently, the anchored field lines tend to flow inward with almost the same speed as the accreting matter (see also Dyda et al. 2013). However, one issue with this picture is that the suppression of turbulence in the surface layers would halt the accretion and consequently there would be no inward drift of field lines.
A common starting point to express the problem is through the induction equation whose $z$-component in the cylindrical coordinates $(r,\phi,z)$ reads
\begin{equation}\label{field-drag-1}
\frac{\partial(rB_z)}{\partial t}=\frac{\partial}{\partial r} \left(  rB_zv-\eta r (\frac{\partial B_r}{\partial z}-\frac{\partial B_z}{\partial r} ) \right),
\end{equation}
where $v=-v_r$ is the radial accretion speed. The first term on the RHS of the above equation indicates the inward advection of the magnetic field. The term $\partial_r B_z$ characterizes the above equation as a diffusion equation. The last term describes the radial diffusion of the magnetic field and is negligible for $B_r \sim B_z$ near the surface. The term $\partial_z B_r$ is troublesome: this term indicates the annihilation of oppositely directed radial components above and below the disk---magnetic reconnection. Note that this term would be absent if the large scale field were vertical inside the disk with no radial component to reconnect.

Lovelace et al. (2009) has claimed to find stationary solutions indicating that a large scale magnetic field does not diffuse away in an accretion disc with large plasma beta, $\beta>1$. In this rather peculiar picture, the flow is found to be radially outward near the mid-plane and radially inward in the parts near the surface. This study found that Prandtl numbers larger than a critical value $\sim 2.7$ can trigger magneto-centrifugal outflows from the surface. For smaller magnetic Prandtl numbers, this work has found electromagnetic outflows instead of jets. However, numerical simulations, in general, do not imply any sign of meridional circulation in discs. Moreover, the form of the induction equation given by eq.(\ref{field-drag-1}) ignores the divergence-free constraint on the magnetic field, $\nabla.{\bf{B}}=0$. Lovelace et al. (2009) also considered a height-dependent magnetic diffusivity $\eta(z)$ that vanishes at the surface; $\eta(z)\propto (1-z^2/h^2)^\zeta$ with a constant $\zeta$. Although the general expectation is that the diffusivity decreases slowly with height but neither turbulence nor MRI shuts down completely near the surface. At most, we may expect an exponential decrease.

\subsection{Coronal Mechanism}\label{SSBeckwith2009}
\begin{figure}
\includegraphics[scale=.45]{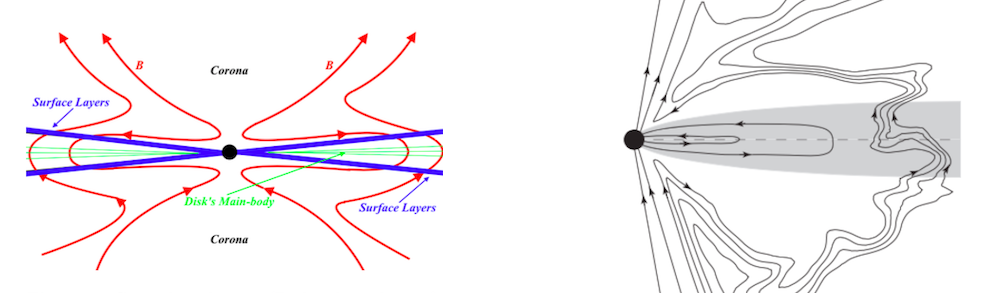}
\centering
\caption {\footnotesize {Right: The illustration of the coronal mechanism proposed by Beckwith et al. (2009). The poloidal field is traced from infinity almost vertically down to the corona where it is advected in, and so sharply bent, toward the central object by efficient inward accretion. The field then moves outward at comparatively lower heights and enters the disc passing through the mid-plane almost vertically mirroring the same structure on the other side of the disk. Reconnection across the mid-plane leads to the formation of magnetic loops that can accrete inward. Left: A simplified picture of the large scale field configuration: the coronal mechanism is similar to the model proposed by Bisnovatyi-Kogan \& Lovelace (2007) except that instead of the surface layers, rapid field advection occurs at higher altitudes in the corona. Also the latter model, unlike the former, does not invoke the reconnection inside the disk as a means to move the field inward. }}
\label{Beckwith2009}
\end{figure}

The "meridional circulation" of Lovelace, Rothstein \& Bisnovatyi-Kogan (2009) that follows from the assumption of the frozen field into the non-turbulent surface layers employed by Bisnovatyi-Kogan \& Lovelace (2007) has been observed in some other numerical works such as Stone \& Norman (1994), Beckwith, Hawley \& Krolik (2009) and, even for protoplanetary discs, Takeuchi \& Lin (2002). On the other hand, some other authors, for example, Fermong, Lyra \& Masset (2011), have argued with a proper consideration of  turbulent and viscous stresses there would be no such a motion. The recent paper by Zhu \& Stone (2017) has also reported a meridional motion commenting that this sort of motion would be possible only in simulations with no net toroidal field. In any case, such an inward motion at the corona and outward motion across the mid-plane (which in general doesn't need non-turbulent surface layers) would lead to a sharp pinch in the field lines near the surface and would probably lead to reconnection. It also would affect our picture since such a structure induces strong magnetic torques near the surface and affects the bending angle; See Fig.(\ref{Beckwith2009}). 

In fact, the "coronal mechanism", proposed by Beckwith et al. (2009), is similar to the "meridional" model found analytically by Lovelace et al. (2009) following the non-turbulent surface assumption employed by Bisnovatyi-Kogan \& Lovelace (2007). However, in the former work the ensuing reconnection in the disc is invoked as a mechanism to advect magnetic flux in the disc: see Fig.(\ref{Beckwith2009}). The reconnection in the corona would not affect flux distribution in the disc whereas the above argument presented in Beckwith et al. (2009) refers to reconnection "inside" the disc. The other distinction is that

\subsection{Axial Symmetry Breaking}\label{SSSpruit2005}

Spruit \& Uzdensky (2005) argued that the turbulent diffusion can be effectively reduced by grouping large scale vertical magnetic fields into small bundles through magnetic flux expulsion (Zeldovich 1957; Parker 1963). In these bundles the fields are assumed to be strong enough to quench the underlying turbulence in order to avoid an outward diffusion (Guan \& Gammie 2009). This flux expulsion also occurs on the solar surface. The result of this mechanism is two-folded: first, it reduces the rate of magnetic reconnection, and second, it makes the loss of angular momentum much more efficient for the patches. The latter is the reason that the patches can accrete inward at a higher rate than that of the outward magnetic diffusion. The concentration of the magnetic flux around the central object leads to a bundle of the field lines. In such a disc, called magnetically arrested disc, this strong magnetized bundle will affect the accretion process (Narayan et al 2003; Spruit \& Uzdensky 2005).

\begin{figure}
\includegraphics[scale=.42]{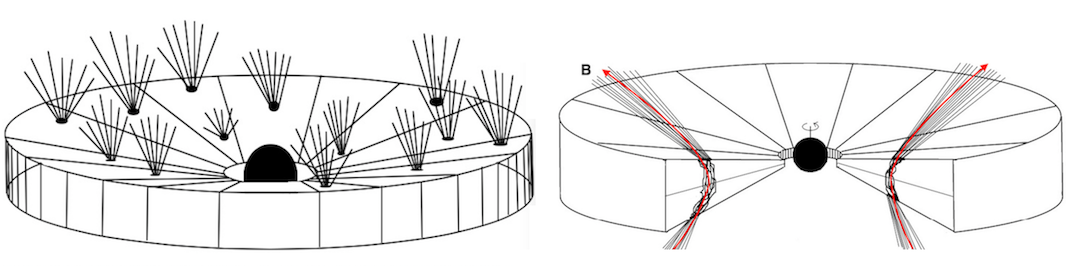}
\centering
\caption {\footnotesize {According to the model suggested by Spruit \& Uzdensky (2005), the large scale magnetic field in the disk is concentrated into multiple patches through magnetic flux expulsion. Magneto-centrifugally driven winds enhance the transport of angular momentum flux in the magnetized patches leading to their rapid inward advection. A loophole in this model is the stability assumption of the magnetized patches in a turbulent medium; see the text.}}
\label{Axisymmetry}
\end{figure}
Fig.(\ref{Patches}) schematically shows some magnetized ``patches'' on a disc. These patches lose angular momentum through efficient outflows that enables them to accrete inward. One may also add to this picture the assumption that the surface layers are weakly turbulent as a result of radiative flux (Bisnovatyi-Kogan \& Lovelace 2007; Jafari \& Vishniac 2017). Magnetized bundles can form in the main body of the disc with their footpoints at the base of the surface layers. Large scale radial field has a negligible mean value on the mid-plane since magnetic buoyancy and turbulent pumping drag the flux tubes up the surface. Hence, the ensemble average of radial field component in a bundle entering the disc through a patch should be negligible across the mid-plane. Also, the diffusion rate in a magnetized bundles turns out to be much smaller than the inward turbulent pumping. Therefore, the magnetized bundles can ``preserve'' their identity for a long enough time to reach the center of disc. Consider a flux bundle with surface area $a^2$. The magnetic tension force is $ a^2  B_rB_z/4\pi$. The balance between radial forces acting on the volume is 
\begin{equation}
{GM_*\delta M\over r^2}-a^2k {B_z^2\over 4\pi}=\delta M{(v_\phi-\delta v_\phi)^2\over r},
\end{equation}
where $\delta M=\rho h a^2$ with $\rho$ being the density. This yields 
\begin{equation}
\delta v_\phi\simeq k {v_A^2\over c},
\end{equation} 
where $v_A$ and $c_s$ are, respectively, the local Alfv\'en and sound speeds inside the bundle. Since, the MRI is suppressed inside the bundle, so $c_s^2/v_A^2\sim 1$. This difference in orbital velocities between the magnetized volumes and their surrouanding matter creates a drag force. We may estimate this by $C\rho \nu a \delta v_\phi$ where $C$ is a constant and $\nu$ the viscosity. The change of angular momentum is given by
\begin{equation}
{dJ_p\over dt}=r {Ck\nu \over c_s} {B_z^2\over 4\pi} a-\gamma r a^2 B_z^2/4\pi.
\end{equation}
The second term in the RHS comes from the magnetic tension $B_\phi B_z/4\pi$ with the parametrization $\gamma=B_\phi/B_z$. The angular momentum is $J_p\simeq \delta M v_K r$ with the Keplerian velocity $v_K$. Thus, the radial drift velocity is
\begin{equation}
v_d={ dJ/dt\over dJ/dr}\simeq  {v_A^2\over c}\Big ({Ck\nu\over4 a c_s}-\gamma\Big).
\end{equation}

We can approximate the diffusion coefficient for the reconnection of radial component in the corona as $D\sim \epsilon \Omega H^2$ where $\epsilon \Omega$ is the rate for the change of magnetic field; $\epsilon\simeq B_r/B_\phi$. This yields an outward drift velocity of order $kD/H$. We get
\begin{equation}\label{Spruit-1}
v_d={ dJ/dt\over dJ/dr}\simeq  {v_A^2\over c_s}\Big ({Ck\eta\over 4a c_s}-\gamma\Big)+{k\epsilon\over 3}c_s,
\end{equation}
where we have substituted magnetic diffusivity $\eta$ for the viscosity. This is exactly the same result that Spruit and Uzdensky (2005) obtained. In the stationary state, according to eq.(\ref{lastk}), $k$ vanishes on the mid-plane, however, so does $\gamma$. Hence, the drift velocity of such a notional flux bundle is zero at the mid-plane. It also vanishes near the surface provided that
\begin{equation}
k\simeq \gamma \Big( {C\eta\over 4 a c_s}+\epsilon {c_s^2\over v_A^2}\Big)^{-1},
\end{equation} 
where all quantities are considered at the surface. 

One question arises about the life-time of the magnetized patches. The diffusion time scale inside a patch of radius $a$ is $\tau_{diff}\sim a^2/\eta_s$. On the other hand, the gradient of magnetic diffusivity inside and outside a patch is of order $\eta_t-\eta_s\sim \eta_t$ that leads to a strong turbulent pumping of ambient field to the patch with a time scale $\tau_p\sim a^2/\eta_t$. We have
\begin{equation}
\tau_{diff}\sim {\eta_t\over \eta_s}\tau_p\gg \tau_p.
\end{equation}
Turbulent diamagnetism pushes the ambient flux into the patch with a time scale much shorter than the diffusion time scale inside the patch.

\subsection{Turbulent Pumping}

The transport of the large scale magnetic field, in inhomogeneously turbulent fluids, by means other than the mean flow of matter is called pumping. One such mechanism is turbulent diamagnetism which results from a gradient in turbulent magnetic diffusivity (Zeldovich 1957; Spitzer 1957). This phenomenon is thought to have negligible effect in the solar convection zone (Spruit 1974). The suppression of turbulence at the surface layers of an accretion disc, with lower optical depth, by the radiative flux (Bisnovatyi-Kogan \& Lovelace 2007) leads to a lower magnetic diffusivity even if the turbulence is not fully suppressed. Magnetic diffusivity is estimated by the turbulent diffusivity $\alpha_{ss} h c_s$ in the disc's main turbulent part. The lower diffusivity at the surface creates a gradient leading to a vertical motion of field lines. This mechanism, known as turbulent pumping or turbulent diamagnetism, has the same effect as the magnetic buoyancy. Both tend to hinder the reconnection of the radial field on the mid-plane which leads to an inefficient outward diffusion. 

Pumping refers to the transport of the large scale magnetic field, in inhomogeneous turbulent fluids, by means other than the mean flow of matter. One such a mechanism is turbulent diamagnetism that results from a gradient in turbulent magnetic diffusivity (Zeldovich 1957; Spitzer 1957). This mechanism is thought to be negligible in the solar convection zone (Spruit 1974). The partial suppression of turbulence at the surface layers of an accretion disc, with lower optical depth, by the radiative flux (Bisnovatyi-Kogan \& Lovelace 2007) leads to a lower magnetic diffusivity even if the turbulence is not fully suppressed. Magnetic diffusivity is estimated by the turbulent diffusivity $\alpha_{ss} h c_s$ in the main turbulent body. The lower diffusivity at the surface creates a gradient leading to a vertical motion of field lines. This mechanism, known as turbulent pumping or turbulent diamagnetism, has the same effect as the magnetic bouyancy. Both tend to hinder the reconnection of the radial field on the mid-plane which leads to an inefficient outward diffusion. Another pumping mechanism is the density pumping or $\nabla \rho$-effect that results from inhomogeneities in density. Drobyshevski (1977) found the effective transport velocity of large scale field, in a two-dimensional density stratified rotating fluid, as $\eta_t \nabla \ln\rho$ where $\rho$ is density and $\eta_t$ turbulent magnetic diffusivity. This is a well-known phenomenon in the solar dynamo theories and can be quite important as an anti-buoyancy effect when fast rotations and strong magnetic fields are present. However, Vainshtein (1978) showed that this effect disappears in three dimensional locally isotropic turbulence (Kichatinov 1991). The role of density pumping in a slowly rotating thin accretion disc threaded by a weak magnetic field may be negligible (Jafari \& Vishniac 2017).

If we take the assumption of non-turbulent surface (Bisnovatyi-Kogan \& Lovelace 2007) seriously then we can estimate a vertical pumping velocity of order 
\begin{equation}
v_p\sim \eta_t/h.
\end{equation}
Although the complete shut-down of the MRI in the surface seems unrealistic, not to mention other hydrodynamic, MHD and plasma instabilities, but this would help to illustrate the possible role of turbulent diamagnetism. In the $\alpha$-prescription (Shakura-Sunyaev 1973), the turbulent diffusivity reads
\begin{equation}\label{t-diffusivity}
\eta_t(r)\simeq \eta_0 r^{1/2},
\end{equation}
where $\eta_0=\alpha_{ss}\sqrt{GM_*} (H/ R)^2$ with Shakura-Sunyaev (1973) constant $\alpha_{ss}$. 
For thin and fully ionized accretion discs, $\alpha_{ss}\sim 0.1-4.0$ and $H/R$ is assumed to be constant. Nevertheless, this is just an approximation since $h$ is in general as function of radius. Assuming that the disc radiates as a black body, the temperature can be shown to have the profile $T\propto r^{-3/8}$, therefore $h/r\propto r^{9/8}$. However, we need to take care of accretional as well as irradiational heating. This gives rise to a temperature profile as $T\propto r^{-1/2}$. Thus, we get $c_s\propto r^{-1/4}$ that leads to $h/r \propto r^{1/4}$. A typical observed accretion rate for young T Tauri stars of the solar mass is $10^{-7} M_{\odot}/yr$, so a typical disk temperature would be $150 K$ at radius $1 AU$. Gammie (1996) used $h/r\propto r^{1/5}$ for T Tauri discs with layered structures.

Numerical simulations suggest values for $\alpha_{ss}$ one order of magnitude smaller than what observations indicate (King et al. 2007). The corresponding radial turbulent pumping velocity is $-\eta_t/r$ which is of the order of the accretion speed. The above expression is the averaged turbulent diffusivity over the height of the disc. In order to recover a $z$-dependence for $\eta_t$, we need a symmetric function whose average over the disc's height is $\eta_0 r^{1/2}$. The function should also be decreasing as $z$ approaches the surface. Lovelace et al. (2009) took a $z$-dependence for the effective diffusivity which essentially has the following form;
\begin{equation}
\eta=\eta_t \Big(1-{z^2\over h^2}\Big)^\zeta,
\end{equation}

where $\zeta$ is a numerical constant. However, this expression yields a zero diffusivity at the surface where we expect a lower, not necessarily zero, magnetic diffusivity. Instead we may consider a fast exponential decline similar to the density profile. The vertical structure of the disk is governed by the equation of hydrostatic equilibrium;
\begin{equation}\label{hydrostatic-accretion-1}
\frac{dP}{dz}=-\rho \frac{GM_*}{(r^2+z^2)} \frac{z}{(r^2+z^2)^{1/2}} \simeq -\rho z \frac{GM_*}{r^3},
\end{equation}
where in the second equation we have used the thin-disk approximation; $h/r \ll 1$. Using $P=c_s^2 \rho$, we get the solution
\begin{equation}\label{hydrostatic-accretion-2}
\rho(r,z)=\rho(r)e^{ -z^2/2h_g^2},
\end{equation}
where $h_g=(c_s^2r^3/GM_*)^{1/2}=c_s/\Omega$ is the gas pressure height scale and $\rho(r)=\rho(r, z=0)$. We may use a similar profile for the diffusivity

\begin{equation}\label{eff-diff}
\eta(r,z)=\eta_t (r) e^{-\xi z^2/2h^2},
\end{equation}
where $\eta_t(r)=\alpha_{ss}c_s h\propto r^{1/2}$ is the turbulent diffusivity on the mid-plane. Taking $\xi=1$, this becomes similar to the choice made by Fleming and Stone (2003) for a layered disc. However, for a thin disc, we assume that $\xi$ is of order, but somehow larger than, unity. The corresponding vertical pumping velocity, $-\partial_z \eta_t$, is $v_p\sim \xi z\eta_t/h^2$ which is of order $\eta_t/h$ as we estimated before. 

The $z$-component of the induction equation in an axisymmetric cylindrical geometry is given by
\begin{equation}
{\partial B_z\over\partial t}\simeq-{\partial E_\phi\over \partial r},
\end{equation}
which vanishes in the stationary state. The integration then gives us the Ohm's law in the azimuthal direction;
 \begin{equation}\label{General-Ohm-Law-2}
\eta({\nabla}\times\textbf{B})_{\phi}=(\textbf{v}\times\textbf{B})_{\phi},
\end{equation}
The RHS of the above expression originates from vertical turbulent pumping, inward accretion, radial turbulent pumping velocity and vertical density pumping. The contributions of the vertical and radial turbulent diamagnetism can be written as
 \begin{eqnarray}
{\cal{E}}_{\phi}^{TD}&=&{\partial \eta\over \partial r} {B}_z-{\partial \eta\over \partial z} {B}_r.
\end{eqnarray}

The contribution of the inward accretion velocity $v_r\sim -\nu/r$ is
 \begin{equation}\label{Buoyancy-3}
{\cal{E}}_{\phi}^A\simeq -v_r {B}_z\simeq \frac{\eta}{r}{B}_z,
\end{equation}
where we assumed a magnetic Prandtl number of order unity; $\nu\simeq \eta$.

The gas density is inhomogeneous over the half thickness of the disc, so a magnetic transport effect appears in the vertical direction which is called density pumping in the solar context (Krivodubskij 2005; Kichatinov 1991; Drobyshevskij 1977). The corresponding transport velocity is given by
\begin{equation}
v_d=\phi {\eta\over \rho}\nabla \rho.
\end{equation}
Here, $\phi$ is a function of angular velociy, $\Omega$, and magnetic field, $B$, and vanishes if both $\Omega$ and $B$ are small. It can be large for rapid rotations and strong magnetic fields in stars but also vanishes for very strong magnetic fields (Rudiger \& Hollerbach 2004). For an accretion disc with a weak magnetic field and slow angular velocity, we can take it as a small constant, $\phi\ll 1$. Another  justification to ignore this effect comes from the work of Vainshtein (1978) who showed that the density pumping is negligible in three dimensional locally isotropic turbulence. Since the density of the disc varies with its height as $\rho(r,z)\propto e^{-z^2/2h^2}$, so we find the downward density pumping velocity as 
\begin{equation}
v_\rho\simeq -\phi \eta {z\over h^2}.
\end{equation}
The corresponding electromotive force is
\begin{equation}\label{rho-0}
{\cal{E}}_{\phi}^{\rho}\sim -\phi {z\over h^2}\eta k B_z.
\end{equation}
Assuming $\phi\ll1$, this is negligible compared with the electromotive force resulted from the upward pumping:
\begin{equation}\label{rho-0}
{\cal{E}}_{\phi}^{\eta}\sim \xi {z\over h^2}\eta k B_z,
\end{equation}
where $k=B_r/B_z$.  The LHS of eq.(\ref{General-Ohm-Law-2}) is 
 \begin{equation}\label{Buoyancy-2}
{\cal{E}}_{\phi}^D=\eta\left( \frac{\partial {B}_r}{\partial z}-\frac{\partial {B}_z}{\partial r} \right).
\end{equation}

Taken together, these lead to the following differential equation:
\begin{equation}\label{eq-k-1}
{\partial(k\eta B_z)\over \partial z}=-\phi {z\over h^2}(k\eta B_z)+{1\over r} {\partial (r\eta B_z)\over \partial r}.
\end{equation}
We know that magnetic diffusivity is a function of radius $\eta\propto r^{1/2}$. Also, in a magnetically arrested disc, we can write a power-law radius-dependence for the vertical magnetic field. For example, Bisnovatyi-Kogan and Lovelace (2007) found $B_z\propto r^{-1}$. Here, we take a general form as $B_z=B_0  r^{-N}$ with $N>0$ and use eq.(\ref{eff-diff}) for magnetic diffusivity to write
\begin{equation}
\eta B_z\simeq\eta_0B_0 r^{-N+1/2} \exp{(-\zeta z^2/2h^2)}, 
\end{equation}
where$B_0$ is a constant. This leads to the following estimate for $k(r,z)$;
\begin{equation}\label{lastk-2-1}
{B_r\over B_z}\simeq \Big( {H\over R}{3-2N\over 2}\Big) {z\over h}e^{(\zeta-\phi)z^2/2h^2}.
\end{equation}
With a negligible $\phi$, this predicts an almost vertical field in the main body of the disc while it retains a radial component comparable to the vertical field at the surface.

\subsection{Magnetic Buoyancy}

Magnetic buoyancy can prevent turbulent mixing of the magnetic field. While the radial magnetic field can be negligible near the mid-plane but it can increase rapidly toward the surface. Suppose the turbulence is sub-sonic, that is, the mass transfer due to the turbulence is negligible. In the vertical direction, the change in the pressure is related to the vertical component of gravitational force. This generates a vertical buoyant force with the acceleration $(\Delta\rho/\rho)g_z$ where $g_z=zGM_*/r^3$ is the gravitational acceleration. The buoyant velocity is 
\begin{equation}
v_B\simeq z\tau_c\Omega^2 \left( \frac{V_A}{c_s}\right)^2\simeq \eta{z\over h^2}.
\end{equation}
The corresponding electric field is
\begin{equation}\label{Buoyancy-0}
{\cal{E}}_{\phi}^{B}\sim {z\over h^2}\eta k B_z,
\end{equation}
where $\eta\simeq V_A^2 \tau_c$ is the turbulent diffusivity and $k=B_r/B_z$.

Combining this result with the Ohm's law, eq.(\ref{General-Ohm-Law-2}), we find
\begin{equation}\label{buoyant-10}
{\partial k\over\partial z}=k{z\over h^2}+{1\over r},
\end{equation}
with the initial condition $k(z=0)=0$. To solve this equation, we consider the limits when $z\rightarrow h$ and $z\rightarrow 0$. The former leads to $k\sim e^{z^2/2h^2}$ near the surface while the latter gives us $k\sim z/r$ near the mid-plane;
\begin{equation}\label{buoyant-11}
k\sim {z\over r}e^{z^2/2h^2}.
\end{equation}

This has the same physical effect of pushing the flux upward to the surface as the turbulent pumping effect; eq.(\ref{lastk-2-1}). In fact, we can combine the effect of magnetic buoyancy with turbulent pumping discussed above to write the Ohm's law as (Jafari \& Vishniac 2017):

\begin{equation}\label{eq-k}
{\partial(k\eta B_z)\over \partial z}=(1-\phi){z\over h^2}(k\eta B_z)+{1\over r} {\partial (r\eta B_z)\over \partial r},
\end{equation}
If we ignore the dependence of $B_z$ on both radius and height and dependence of diffusivity on height then we recover eq.(\ref{buoyant-11}). However, in general we write the vertical magnetic field as $B_z=B_0  r^{-N}$ with $N>0$ and $B_0$ being a constant and take a variable diffusivity as e.g., suggested by eq.(\ref{eff-diff}). We find
\begin{equation}
\eta B_z\simeq\eta_0B_0 r^{-N+1/2} \exp{(-\zeta z^2/2h^2)}, 
\end{equation}
This leads to the following expression for $k(r,z)$:

\begin{equation}\label{lastk}
k(r,z)\simeq\Big[({3\over 2}-N)\sqrt{{\pi\over 2(\zeta-\phi+1)}} {h\over r}  erf\Big({z\over h} \sqrt{{\zeta-\phi+1\over 2} } \Big)+k_0(r)\Big]e^{(\zeta-\phi+1)z^2/2h^2},
\end{equation}
where we choose $k_0(r,0)=0$ as the constant of integration with respect to $z$ since the inclination angle, supposedly, vanishes on the mid-plane. Also, $z/h$ is of order unity, and we can approximate the error function as $erf(x)\simeq 2x/\sqrt{\pi}$. We find
\begin{equation}\label{lastk-2}
{B_r\over B_z}\simeq \Big( {H\over R}{3-2N\over 2}\Big) {z\over h}e^{(\zeta-\phi+1)z^2/2h^2},
\end{equation}
where we may assume $H/R=h/r$ is a constant despite that it slightly depends on the radius. Note that the ratio $B_r/B_z$ is negligibly small for all values of $z$ up to few scale heights, but then, it quickly approaches unity. In a standard $\alpha$-disc, the stress shear $w_{r\phi}=\rho V_A^2$ is given by $w_{r\phi}=\alpha_{ss} \rho c_s^2$, therefore the turbulent magnetic diffusivity is $\eta_t=B^2/8\pi \rho\Omega$ with $\rho\propto r^{-15/8}$. On the other hand, $\eta_t=\alpha_{ss}hc_s\propto r^{1/2}$. It follows then $B\propto r^{-23/16}$, so $N\simeq 23/16$. As all our estimates rely on scaling laws, rather than rigorous calculations, the final result can be written in a more convenient form as 

\begin{equation}\label{Final}
{B_r\over B_z}\simeq {z\over r}e^{z^2/h^2}.
\end{equation}

Accretion discs, even thin ones, have vertical structure. The poloidal field threading the disc is in fact mostly vertical in the disc except at its surface layers where it deviates from the vertical giving rise to an appreciable radial component which is required for efficient outflows. This is in agreement with the fact that the ratio of radial to vertical components is of order unity (Bisnovatyi-Kogan \& Blinnikov 1972; Ustyugova et al. 1999; Bisnovatyi-Kogan \& Lovelace 2007). Magnetic buoyancy, operating in the main body, and turbulent diamagnetism, acting near the surface, both impede the turbulent reconnection and outward diffusion by reducing the magnitude of the radial field in the disc's main body. 

\subsection{Vertical Drift Model}\label{SSJafari2017}

Mechanisms discussed in the previous sections (buoyancy, turbulent diamagnetism and pumping) can be combined through the standard MHD equations to calculate the net vertical transport velocity for the flux tubes. The net vertical drift velocity will reduce the radial component near the mid-plane while it can lead to a large bending angle at few scale heights. Jafari and Vishniac (2018) found the following vertical drift velocity in anisotropic turbulence 
\begin{equation}\label{buoyant-v}
\langle v_z\rangle\simeq \tau{\langle b_\phi^2\rangle\over 4\pi \rho_0 \gamma L_p},
\end{equation}
where $L_p=p/\partial_z p$ is the scale height of pressure $p$; $ b_\phi$ and $v_z$ are respectively the azimuthal small scale magnetic field and vertical velocity field; $\rho_0$ is the density at the mid-plane; $\tau$ is the eddy turn-over time and $\gamma=c_p/c_v$ is the specific hear capacity ratio. In order to find the bending angle one may appeal to a general consideration independent of any particular model prescribed for the viscosity or the magnetic diffusivity. In the steady state, the total electric field should vanish so we balance the electric fields induced by the vertical buoyant velocity and radial accretion velocity with the resistive diffusion through the Ohm's law.  In the azimuthal direction;

\begin{figure}
\includegraphics[scale=.35]{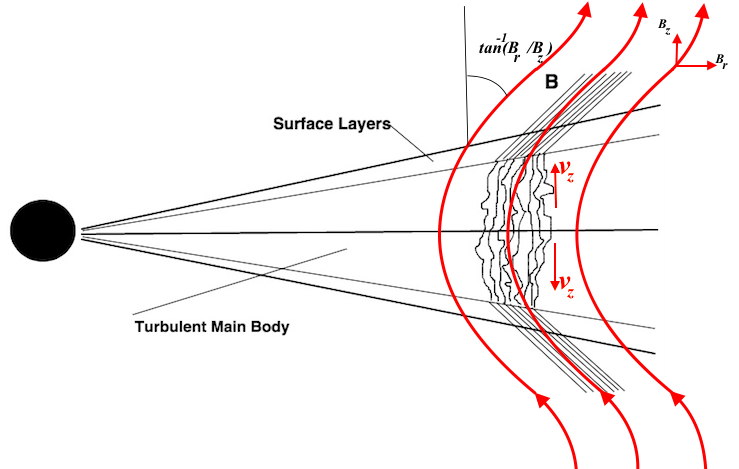}
\centering
\caption {\footnotesize {Cross section of a turbulent thin disc threaded by a large scale poloidal magnetic field. The magnetic buoyancy, in the subsonic turbulence, and turbulent pumping together push the field lines towards the surface layers of the disc decreasing the efficiency of radial reconnection across the mid-plane. One implication is the dissipation timescale is independent of the disc's height (Jafari \& Vishniac 2018).}}
\label{supp-diff}
\end{figure}

\begin{equation}
\langle v_z\rangle B_r-{V_0}_r B_z=\langle v_z^2\rangle \tau {\partial B_r\over \partial z}-\langle v_r^2\rangle \tau {\partial B_z\over \partial r},
\end{equation}
where ${V_0}_r=-(3\nu/2r)=-|{V_0}_r|$ is the radial accretion velocity at the mid-plane. We can estimate the vertical drift with the buoyant velocity given by the second term of eq.(\ref{buoyant-v}), ignoring the first term since we assume $\langle b_\phi^2\rangle \ll \langle v_z^2\rangle$ as discussed before. So we get $\langle v_z\rangle\simeq \tau(\langle b_\phi^2\rangle/4\pi \rho_0 \gamma L_p)$. This leads to
\begin{equation}
-B_r {\langle b_\phi^2 \rangle \over 4\pi \rho_0 \gamma L_p} \tau\simeq \langle v_z^2\rangle \tau {\partial B_r\over \partial z}-\langle v_r^2\rangle \tau {\partial B_z\over \partial r}-|{V_0}_r| B_z.
\end{equation}
Hence, we find

\begin{equation}\label{Pgrad}
-{1\over 4\pi\gamma}{\partial\ln{P}\over \partial z}={\rho \langle v_z^2 \rangle  \over  \langle b_\phi^2\rangle}  {\partial \ln{B_r}\over \partial z}-{\rho\langle v_r^2\rangle \over \langle b_\phi^2\rangle } {1\over B_r}{\partial B_z\over \partial r}+{\rho\over\tau\langle  b_\phi^2\rangle }|{V_0}_r| \Big({B_z\over B_r}  \Big).
\end{equation}
On the other hand, at the mid-plane, we can also write $\langle v_z^2 \rangle \tau (\partial B_r/\partial z)=|{V_0}_r |B_z$. Thus we can estimate the radial field at very small heights as $B_r\simeq zB_z(|{V}_r |+\langle v_r^2\rangle \tau/\langle v_z^2\rangle \tau  )_0$. Substituting this in the above equation we find

\begin{equation}\label{general2}
{B_r\over B_z}\simeq z \Big( {|{V_0}_r|-\langle v_r^2\rangle\tau\partial_r \ln B_z\over \tau \langle v_z^2\rangle} \Big)\Big( {P\over P_0}\Big)^{-{\langle b_\phi^2\rangle\over4\pi\gamma\rho\langle v_z^2\rangle}}.
\label{bent}
\end{equation}

Let us write $\alpha c_s^2 \simeq \langle b_\phi^2\rangle /4\pi\rho$, where $\alpha$ is a constant which parametrizes the local turbulence in terms of the pressure and scales with the Shakura-Sunyaev (1973) parameter $\alpha_{SS}$ but typically is larger. Assuming $\Omega\tau\simeq 1$, we can take $\alpha c_s^2/\Omega\tau \langle v_z^2\rangle$ approximately equal to $ \langle b_\phi^2\rangle/4\pi\rho \langle v_z^2 \rangle$. This is the ratio of effective buoyant term to turbulent mixing that is of the same order of the exponent in the last brackets. This is also approximately equal to the ratio of pressure over magnetic scale heights $L_p/L_B$ (see equation (\ref{BC1}) below). Numerical work suggests a value roughly equal to $5$ for this coefficient (see Blackman et al. 2008; Jafari \& Vishniac 2018). We can write

\begin{equation}\label{N}
N= {\alpha\over \Omega\tau}{ c_s^2\over\langle v_z^2\rangle}  \simeq {\langle b_\phi^2\rangle\over 4\pi\gamma\rho \langle v_z^2\rangle  },
\end{equation}
with the assumption that $N$ does not vary much with height or radius. Consider the velocity competition between advection and diffusion of the large scale field, which appears as the velocity difference $|{V_0}_r|-\langle v_r^2\rangle\tau\partial_r \ln B_z=V_{acc}-V_{diff}$ at the mid-plane in the above expression. We have

\begin{equation}\label{general2-2}
{B_r\over B_z}\simeq z   {|{V_0}_r|\over \tau \langle v_z^2\rangle} \epsilon(r) \Big( {P\over P_0}\Big)^{-N},
\end{equation}

where $\epsilon(r)$ is a function of the radius. The outward radial diffusion flattens an otherwise unreasonably large bending angle at the outer radii. Assuming a large bending angle, a simple ansatz for the global structure is to take $|B| \sim B_r\propto r^{-2}$, and let $B_z\propto r^{-n}$ for some $n>0$ so that $B_r/B_z\propto r^{n-2}$. Equation (\ref{general2-2}) implies $B_r/B_z\propto \epsilon(r) r^{3N/2} $ where $\epsilon(r)=(r'/r)^b$ with $b=3N/2-n+2$ and $0<r'\ll R$. In fact, $\epsilon(r)$ accounts for the behavior of $V_{acc}-V_{diff}$ at radii larger than $r_c$. At these radii we expect $r'/r\ll 1$, i.e. the advection of the field will almost balance with outward diffusion. The field is concentrated at small radii without having an unrealistically steep radial dependence. The normalization of $\epsilon(r)$ depends on the global structure of the magnetic field, in particular the radius $r'$ where the balance between diffusion and inward advection breaks down.

\begin{figure}[h]
\includegraphics[scale=.4]{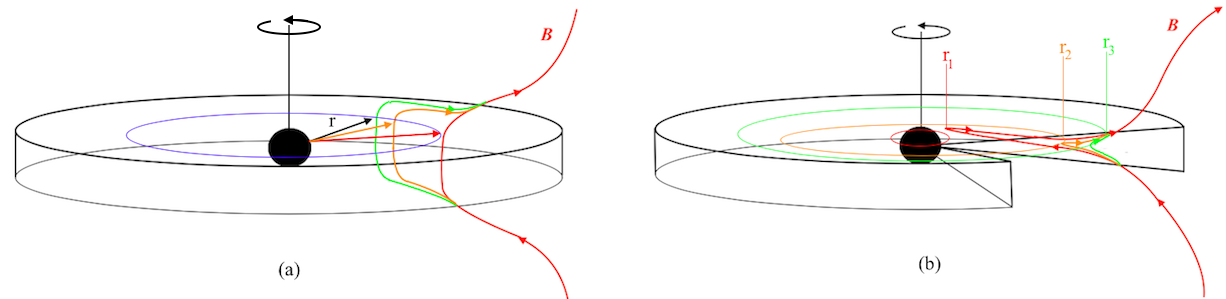}
\centering
\caption {\footnotesize {The effect of buoyancy in a thin accretion disk. (a) A vertical field line (red curve) threads the disk only at one radius $r$ so differential rotation would not affect it. Instead only rotation would stretch the field line producing an azimuthal, but not any radial, component  across the mid-plane. Green and orange curves represent the later evolution of the original field line at the same radius $r$. (b) A tilted magnetic field line, on the other hand, has already a radial component threading the disk over a range of radii so differential rotation would stretch it differently at different radii. Red, green and orange curves here show how the differential rotation affects the original tilted field line differently at different radii $r_1<r_2<r_3$. This brings the oppositely directed field lines across the mid-plane to a close contact which is the ideal configuration for magnetic reconnection. Our MHD treatment indicates that the configuration shown at (b) is not realistic. Diffusion and magnetic buoyancy would lead to a vertical configuration for the field. This in turn reduces the reconnection rate inside the disk. The magnetic field gains a large tilt only at few height scales above the disk. (Illustration from Jafari \& Vishniac 2018)}}
\label{diff-rot}
\end{figure}
To estimate the total bending we need to impose a limiting condition on the  pressure in equation ({\ref{general2-2}). One obvious choice would be to pick the transition from gas pressure dominated to magnetic pressure dominated. However, at somewhat lower heights the gradient of the magnetic pressure will dominate over the opposing gradient of the gas pressure, leading to the launch of an outflow. Assuming a} large bending angle at this point means a magnetic pressure $\sim(B_r^2+B_\phi^2)/8\pi$. So we set the minimum pressure condition in equation (\ref{general2-2}) as
\begin{equation}\label{BC1}
P_{min}\simeq{L_p\over L_B}{B^2\over8\pi}\simeq {\alpha \over \Omega \tau}{c_s^2\over \langle v_z^2\rangle}{ B_\phi^2+B_r^2\over8\pi},
\end{equation}
where $L_B$ is the magnetic pressure scale height. 

Imposing a limiting condition on the  pressure in equation ({\ref{general2-2}) above the disk raises a question about the amount of magnetic field concentration which increases towards the smaller radii. 
Using equation (\ref{N}), the condition for the minimum pressure, given by equation (\ref{BC1}), can be written as
\begin{equation}\label{zap}
P_{min}=N\Big(1+(\Omega\tau_p)^2\Big){B_r^2\over 8\pi},
\end{equation}
where $\tau_p$ is the persistence time in generating the toroidal field from the radial component; $B_\phi\simeq (\Omega \tau_p)B_r$. It is easy to show (see Jafari \& Vishniac for details) that $\Omega\tau_p$ is of order $\alpha^{-1}$  and write $B_r\simeq \alpha B_\phi$. Substituting this result into equation (\ref{zap}), we see that the azimuthal component of the large scale magnetic field will dominate over the other components and therefore:

\begin{equation}\label{finalP}
P_{min}\simeq N {B_r^2\over 8\pi \alpha^2}. 
\end{equation}

We can use any model to illustrate the above results. For example, using the $\alpha$-model (Shakura \& Sunyaev 1973), we may write viscosity as $\nu=\alpha_{SS} c_s h$ in terms of the sound speed $c_s$, the disk's height $h$ and a roughly constant parameter $\alpha_{SS}$. The accretion speed near the mid-plane is then estimated as $|{V_0}_r|\simeq \alpha_{SS} c_s^2/r\Omega$ where $\Omega$ is the angular velocity. Thus, in this model, we estimate the bending of the large scale magnetic field as
 \begin{equation}\label{general4}
{B_r\over B_z}\simeq \epsilon(r) {z\over r} N \Big( {P_0\over P_{min}}\Big)^{N}.
\end{equation}

Equation (\ref{general4}) differs from the result of Lubow et al. (1994) in two ways. The factor $(P_0/P_{min})^N\gg 1$, which is due to the increased efficiency of magnetic buoyancy with height, increases the bending angle very fast near the surface. This effect can give rise to very large bending increasing the importance of radial diffusion which will lead to $\epsilon(r)\ll 1$, partly offsetting the effect of buoyancy. Without accounting for these two effects we would recover a bending angle of several times $h/ r$, consistent with the earlier results.

\subsection{Critical Radius}

The advection concentrates the magnetic flux and compresses the field lines towards the smaller radii. At some radius, the ratio $B_r/B_z$ becomes of order unity. At this radius, the solid angle subtended by the field lines encloses a large fraction of the space above the disk therefore inward this radius the field becomes very strong and stops being efficiently advected anymore. Inward this radius, one may roughly estimate the vertical field as $B_z\sim B_{ext}(R/r)^2$ where $R$ is the outer radius. However, for the larger radii, the radial field component is larger than the vertical component as a result of efficient bending. As a crude estimate, we may write
\begin{equation}\label{concentration}
B_r\simeq B_{ext}\Big({R\over r}\Big)^2.
\end{equation}
Assuming that the bending angle $i=\tan^{-1}(B_r/B_z)$ is given by equation (\ref{general2}), the poloidal field can become dynamically important affecting the structure of the disk. This can happen when the flux of angular momentum carried by the outflows, mediated by the vertical field lines, becomes comparable to or larger than the momentum flux transported through the thin disk (see e.g., Dyda et al. 2013). We also note that the angular momentum loss comes from regions at one or few scale heights from the mid-plane. This can lead to infall which prevents further bending and replaces the $P_{min}$ criterion given by equation (\ref{finalP}). It also implies the presence of dissipation near the photosphere or beyond.

Let us compare the angular momentum loss through the outflows, mediated by the large scale field, to its internal transport through the disk. In an annulus of the radial width $\Delta r$, the angular momentum flux transported radially outward through the disk is $ {\cal{L}}_d=2\pi r h \Delta r(r\alpha_{SS} P_0)  $. The flux carried out by the outflows, on the other hand, is given by ${\cal{L}}_w=4\pi r^2 \Delta r  (rB_z B_\phi/4\pi)$; see Fig.(\ref{wind}). Since $B_\phi\simeq B_r/\alpha$ (see the argument after equation (\ref{zap})), the latter can be written as ${\cal{L}}_w=4\pi r^2 \Delta r  (rB_z B_r/4\pi\alpha)$. At some critical radius $r_c$, ${\cal{L}}_w\simeq{\cal{L}}_d$ and we have
\begin{figure}
\includegraphics[scale=.42]{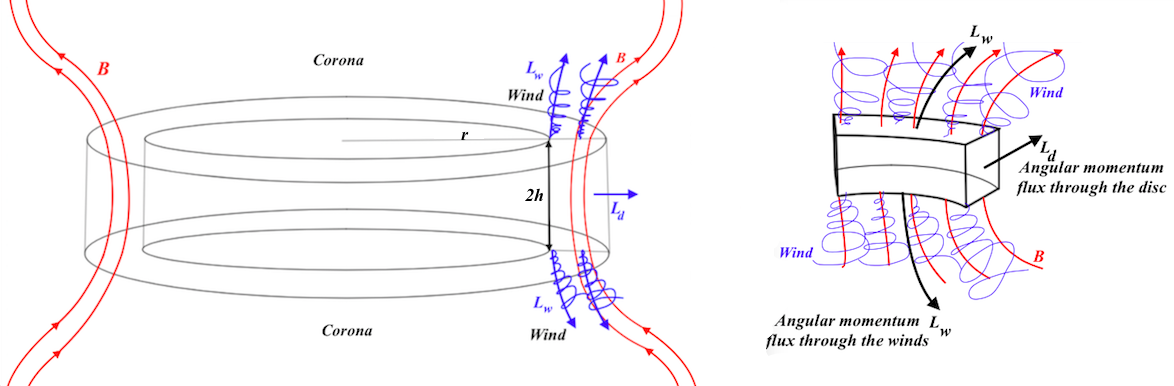}
\centering
\caption {\footnotesize {Angular momentum loss in an annulus of height $h$ at radius $r$ of a thin disk. Angular momentum is lost through both the winds, $L_w$ mediated by the poloidal field $\bf{B}$, and  the disk, $L_d$ mediated by the MRI-turbulent viscosity $\nu=\alpha c_s h$. As the external field $B_{ext}$, trapped by the matter at the outer radii $\sim R$, is advected inward, it becomes concentrated at the inner radii; $B_r\sim B_{ext} (R/r)^2$. Equation (\ref{criterion}) estimates the critical radius inside which the field becomes dynamically important since it can mediate an angular momentum transport, through winds, greater than what is transported through the disk via turbulent viscosity. (Illustration from Jafari \& Vishniac 2018)}}
\label{wind}
\end{figure}

\begin{equation}\label{outer}
\alpha \alpha_{SS}{h\over r_c}\simeq4\Big({ B_rB_z\over 8\pi P_0}\Big).
 \end{equation}
 
 This result actually shows that the field can still be largely bent at the critical radius $r_c$, where the outflows begin to carry more angular momentum flux than what the disk transports internally, provided that $r'$ is not much different than $r_c$. In fact, combining the above condition with the minimum pressure condition given by equation (\ref{finalP}), the bending angle given by equation (\ref{general4}) becomes

  \begin{equation}\label{critical-bending}
 \Big({B_r\over B_z}\Big)_{r=r_c}\simeq \Big( {\alpha\over\alpha_{SS}}\Big)^{N/(N+1)}\Big( {r'\over r_c} \Big)^{b/(N+1)}\Big( N{h\over r_c} \Big)^{(1-N)/(N+1)}.
 \end{equation}
The first and third factors on the RHS are large, and the second factor has a small exponent. Thus, for $r'$ not much smaller than $r_c$, we find that $B_r/B_z$ is large. The poloidal field becomes dynamically important at the critical radius $r_c$, where we still  have a large bending $B_r/B_z\gg 1$. Efficient outflows are expected to prevail inward the critical radius, where the disk transitions to an outflow-dominated regime and accretion speeds up, down to some smaller radius at which $B_r/B_z$ becomes equal or less than $\pi/6$ (Blandford \& Payne 1982) and therefore the efficient outflows cannot launch anymore. We also expect the critical radius to lie outward the radius $r_1$ where $B_r/B_z=1$. The transition to the outflow-dominated state from an accretion-dominated regime at the larger radii inward the critical radius, $r_c$, will probably increase the value of $r_1$ which could be smaller otherwise. Advecting inward toward even smaller radii, the field will eventually become almost vertical, $B_r/B_z\sim 0$, at some inner radius $r_0< r_c$. 
 
 In order to get an estimate of $r_c$, one may combine the condition for efficient outflows, equation (\ref{outer}), with the expression for the bending angle, equation (\ref{general4}), and the minimum pressure condition, equations (\ref{finalP}) along with (\ref{concentration}). We find the critical radius as

\begin{equation}\label{criterion}
 {r_{c}\over R}\simeq  \Big( {r'\over R} \Big)^{2b\over 2b-3(N+1)}    \Big({H\over R} \Big)^{4\over 2b-3(N+1)}\Big(\alpha {\dot M \Omega_R \over  H B_{ext}^2}    \Big)^{2(N+1)\over2b-3(N+1)}\Big( {\alpha\over\alpha_{SS}} \Big)^{2N\over 2b-3(N+1)}N^{{2(1-N)\over 2b-3(N+1)}},
\end{equation}

where $\Omega_R=(GM/R^3)^{1/2}$ is the angular velocity at the outermost radius $R$ where the thickness is $H$ with $H/R\sim constant$. We have also used the mass accretion rate $\dot M= 2\pi v_r \rho h r$ to estimate the mid-plane pressure as $P_0=\rho_0 c_s^2\simeq \dot M \Omega/2\pi\alpha_{SS} h$. This can be written as $P_0\simeq (\dot M\Omega_R/2\pi\alpha_{SS} H) (R/r)^{5/2}$. Assuming $N\sim 5$ and $r'\sim r_c$, we find
\begin{equation}\label{criterion+2}
 {r_{c}\over R}\simeq \Big({\alpha_{SS}\over\alpha} \Big)^{5/9}\Big[ \Big({H\over R}\Big)^{2/3} { B_{ext}^2R \over \alpha \dot M\Omega_R}\Big]^{2/3} .
\end{equation}

The bending angle $B_r/B_z$ is sensitive to the exponent $N$ for which we have used only a rough estimate, $N=5$. With this value for $N$, if we substitute the typical values for SS Cygni (see Schreiber et al. 2003; Bitner et al. 2007 and Miller-Jones et al. 2013),  as an example, during outburst then
we get $r_c\sim 10^{-3} R$. In other words, the critical radius is inside the inner edge of the disk.  During quiescence this radius will move outward and may lie inside the disk. For analogous black hole systems the critical radius will always lie outside the event horizon and inside the disk.

Using our model, we can hardly say anything about the physics inward of the critical radius. At $ r\lesssim r_c$ , the winds may eject a significant mass flux as well as angular momentum so the pressure and density can drop below what we would otherwise expect in an $\alpha$-disk. If so, the maximum bending angle may decrease rapidly until  the field becomes almost vertical. 

 \begin{figure}
\includegraphics[scale=.65]{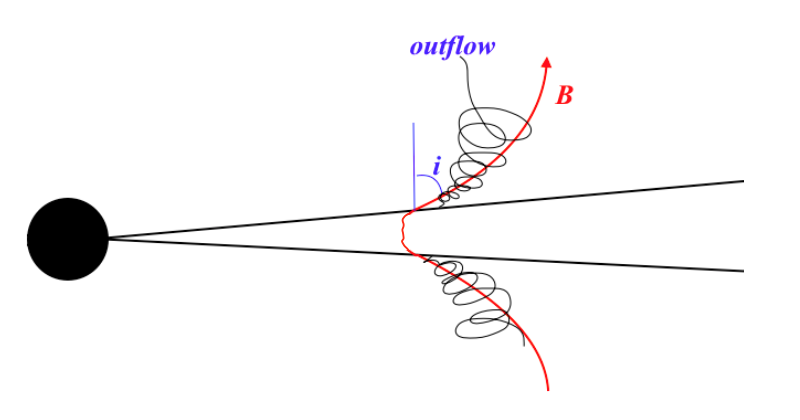}
\centering
\caption {\footnotesize {At the critical radius $r_c$, given by equation (\ref{criterion}), the angular momentum loss through the outflows mediated by the large scale magnetic field becomes greater than its internal transport through the disk. Inward $r_c$ we encounter a more complicated situation which cannot be described using our model. The magnetic field, which still has a large bending angle at $r_c$ as equation (\ref{critical-bending}) indicates, may continue to be advected inward while the bending angle decreases and at some smaller radius we get $B_r/B_z\simeq1$. At this radius, the solid angle subtended by the field lines still would enclose a large fraction of the space above the disk, where the external field lines reside, so inward this radius the field becomes even stronger while it stops being efficiently advected. The field may pile up more inward until it becomes almost straight at even smaller radii. (Illustration from Jafari \& Vishniac 2018) }}
\label{Binside}
\end{figure}

\section{Summary}

Magnetic fields threading accretion disks play key roles in the evolution of the disks and their jets. Numerical simulations confirm that the non-linear regime of the MRI is influenced by the presence of a net vertical field (Fleming et al. 2000). The flux of angular momentum and energy moving out along the magnetic field lines also depend on the vertical field $B_z$ threading the disk (Bai \& Stone 2013). In the presence of a large scale magnetic field, a large bending angle $i=\tan^{-1}(B_r/B_z) \leq 30^\circ$ is required to launch outflows and jets (Blandford \& Payne 1982). On the other hand, magnetic reconnection across the mid-plane of the disk will annihilate the large scale field unless it is almost vertical in the main body of the disk (see e.g., Lubow et al. 1994a, Spruit \& Uzdensky 2005; Lovelace et al. 2009a; 2009b; Dyda et al. 2013). 

Different mechanisms have been proposed for the inward advection of large scale magnetic fields in thin disks. Spruit and Uzdensky (2005) suggested that the large scale field became concentrated in highly magnetized patches inside which the MRI would have ceased as a result of strong magnetic fields. However, it remains unclear, in this picture, how these magnetized patches can keep their geometry, over the accretion time scale, in a turbulent disk susceptible to different hydrodynamic as well as magnetohydrodynamic and plasma instabilities. In another picture, Bisnovatyi-Kogan and Lovelace (2007) suggested that the field could become frozen into, and advected with, the non-turbulent surface layers of the disk. Nevertheless, even though strong magnetic fields and radiation can in principle affect the turbulence at outer layers but a complete shut-down of the turbulence seems unphysical because instabilities such as Parker instability can still make these layers turbulent. Beckwith et al. (2009) suggested that the poloidal field is traced from infinity almost vertically down to the corona of the accretion disk where it is advected in and therefore bent toward the central object by efficient inward accretion. The field then moves outward at comparatively lower heights and enters the disk passing through the mid-plane almost vertically mirroring the same structure on the other side of the disk. Reconnection across the disk leads to the formation of magnetic loops inside the disk that can move inward as the authors argued. This picture differs from what generally is seen in typical numerical simulations and requires further numerical evidence.

Jafari and Vishniac (2018) considered different mechanisms which may affect the large scale field in a thin disk. The showed that the large scale field threading a thin accretion disk will actually remains almost vertical in the main body of the disk while, at few scale heights, it would become largely bent providing the necessary condition for launching outflows. This picture relies on magnetic buoyancy and turbulent pumping although the former plays a much more important role than the latter. It turns out that in fact thin disks can indeed support large bending angles required by Blandford-Payne (1982)  mechanism. This approach predicts a negligible radial field inside the disk, which reduces the reconnection rate, and a very large radial field near the surface few scale heights above the mid-plane, which is essential to efficient outflows. Efficient outflows carrying angular momentum flux, larger than what is internally carried through the disk, prevail at a critical radius larger than the radius where the poloidal field becomes almost vertical. The critical radius $r_c$ would roughly lie at the very inner parts of the disk, e.g., at $\lesssim 10^{-4}R$ for a disk of outer radius $R$ around a solar-mass black hole. In any case, what really happens inward this radius remains unsettled.

\section{references}
Bai, X.-N., \& Stone, J. M. 2011, ApJ, 736, 144\\
Bai, X-N, Stone, \& J. M. 2013, ApJ, 767, 30\\
Balbus, S. A. \& Hawley, J. F. 1991, ApJ, 376, 214\\
Balbus, S. A. 2014, MNRASL, 444, L54âL57\\
Beckwith, K., Hawley, J. F., \& Krolik, J. H. 2008, ApJ, 678, 1180\\
Beckwith, K., Hawley, J. F., Krolik, J. H. 2009,  ApJ, 707, 428 \\
Blandford, R. D., \& Znajek, R. L. 1977, MNRAS, 179, 433\\
Blandford, R. D., \& Payne, D. G. 1982, MNRAS, 199, 883\\
Blandford, R. D. 1993, in Burgarella, D., Livio, M., O'Dea, C. P., eds, Astrophysical Jets. Cambridge Univ. Press, Cambridge\\
Bisnovatyi-Kogan, G. S., \& Blinnikov, S. I. 1972, Ap\&SS, 19, 93\\
Bisnovatyi-Kogan, G. S. \& Ruzmaikin, A. A. 1976, Astroph. \& Space Science, 42, 401\\
Bisnovatyi-Kogan, G. S., \& Lovelace, R. V. E. 2007, ApJ, 667, L167-L169\\
Bisnovatyi-Kogan, G. S., \& Lovelace, R. V. E. 2012, ApJ, 750, 109\\
Brandenburg, A., Nordlund, A., Stein, R. F., Torkelsson, U. 1995, ApJ, 446, 741\\
Burm, H., Kuperus, M. 1988, A\&A, 192, 165\\
Camenzind, M. 1994, in Duschl W. J. et al., eds, Theory of Accretion Discs--2. Kluwer, Dordrecht, p. 313\\
Campbell, C. G. 1987, MNRAS, 229, 405\\
Cao, X, \& Spruit, H. C., 2013, ApJ, 765, 149\\
Cattaneo F., \& Vainshtein S. I., 1991, ApJ, 376, L21\\
Chandrasekhar, S. 1961, Hydrodynamic and Hydromagnetic Stability (New
York: Dover)\\
Coroniti, F. V., 1992, ApJ, 244, 537\\
Davis, S. W., Stone, J. M., \& Pessah, M. E. 2010, ApJ, 713, 52\\
De Villiers, J., \& Hawley, J. F. 2003, ApJ, 589, 458\\
Dursi, L. J., \& Pfrommer, C. 2008, ApJ, 677, 993\\
Drobyshevskij E. M. 1977, Ap\&SS 46, 41\\
Dyda, S, Lovelace, R. V. E., Ustyugova, G. V., Lii, P. S., Romanova, M. M.,\& Koldoba, A. V. 2013, arXiv:1212.0468v3 \\
Fendt C., Camenzind M., Appl S., 1995, A\&A, 300, 791\\
Fendt, C. \& Zinnecker, H. 1998, Astron. Astrophys. 334, 750\\
Flaig, M., Kley, W., \& Kissmann, R. 2010, MNRAS, 409, 1297\\
Fleming, T.,\& Stone, J. M. 2003, ApJ, 585, 908\\
Fricsh, U., Pouquet, A., L\'eorat, J., \& Mazur, A. 1975, J. Fluid Mech., 68, 769\\
Fromang, S., \& Nelson, R. P. 2006, A\&A, 457, 343\\
Fromang, S., \& Stone, J. M. 2009, A\&A, 507, 19\\
Gabov, A. S., Sokoloff, D. D., Shukurov, A. 2001, Dynamo and Dynamics, a Mathematical Challenge
NATO Science Series Volume 26, 233\\
Gammie, C. F. 1996, ApJ, 457, 355\\
Guan, X., \& Gammie, C. F. 2009, ApJ, 697, 1901\\
Hawley, J. F. 2001, ApJ, 554, 534\\
Hawley, J. F., Gammie, C. F., \& Balbus, S. A. 1995a, ApJ, 342, 208\\
Hawley, J. F., Gammie, C. F., \& Balbus, S. A. 1995b, ApJ, 440, 742\\
Hawley, J. F., Gammie, C. F., \& Balbus, S. A. 1996, ApJ, 464, 690\\
Hayashi, C. 1981, Prog. Theor. Phys., 70, 35\\
Hirose, S., Krolik, J. H., \& Stone, J. M. 2006, ApJ, 640, 901\\
Igumenshchev, I. V., Narayan, R., \& Abramowicz, M. A. 2003, ApJ,
592, 1042\\
Jafari, A., \& Vishniac, E. 2018, ApJ, 854, 1\\
Jokipii, J. R. 1991, in Sun in Time, ed. C. P. Sonette, M. S. Giampapa, \& M. S. Matthews (Tucson: Univ. Arizona Press), 205\\
Kichatinov, L. L. 1988, Astron. Nachr. 309, 197\\
Kichatinov, L. L. 1991, Astron. Astrophys. 243, 483\\
King, A. R., Pringle, J. E. , and Livio, M. 2007, MNRAS 376 (4), 1740\\
King, A. L., Miller, J. M., Bietenholz, M., Gultekin, K., Reynolds, M., Mioduszewski, A., Rupen, M., and Bartel, M. 2015, Astrophys. J. Lett., 799:L8\\
Konigl, A. 1989, ApJ, 342, 208\\
Krasnopolsky, R., Li, Z.-Y., \& Blandford, R. D. 1999, ApJ, 526, 631\\
Krivodubskij, V. N. 2005,  Astron. Nachr. / AN 326, No. 1, 61\\
Livio, M. 1997, in ASP Conf. Ser. 121, Accretion Phenomena and Related
OutÑows, ed. D. T. Wickramasinghe, G. V. Bicknell, \& L. Ferrario (San
Francisco : ASP), 845\\
Lesur, G., \& Longaretti, P.-Y. 2009, A\&A, 504, 309\\
Lovelace, R.V.E., Romanova, M.M., \& Newman, W.I. 1994, ApJ, 437, 136\\
Lovelace, R.V.E., Newman, W.I., \& Romanova, M.M. 1997, ApJ, 484, 628\\
Lovelace, R.V.E., Bisnovatyi-Kogan, G. S., Rothstein, D. M. 2009, Nonlin. Processes Geophys., 16, 77\\
Lovelace, R. V. E., Rothstein, D. M.,\& Bisnovatyi-Kogan, G. S. 2009, ApJ, 701, 885\\
Lubow S. H., Papaloizou J. C. B., \& Pringle J. E. 1994a, MNRAS, 267, 235\\
Lubow S. H., Papaloizou J. C. B., \& Pringle J. E. 1994b, MNRAS, 268, 1010\\
Lyutikov, M. 2006, Mon. Not. R. Astron. Soc. 373, 73\\
Matsumoto, R., \& Tajima, T. 1995, ApJ, 445, 767\\
McComas, D. J., Alexashov, D., Bzowski, M,  Fahr, H., Heerikhuisen, 5 J., Izmodenov, V., Lee, M. A.,  Mobius, E., Pogorelov, N., Schwadron, N. A., Zank, G. P. 2012, Science, 336, 6086, 1291\\
Miller, K. A., \& Stone, J. M. 2000, ApJ, 534, 398
McKinney, J. C., \& Blandford, R. D. 2009, MNRAS, 394, L126\\
McKinney, J. C., \& Gammie, C. F. 2004, ApJ, 611, 977\\
Narayan, R., Igumenshchev, I. V., \& Abramowicz, M. A. 2003, PASJ, 55, L69\\
Naso, L., Klu\'zniak, W., Miller,  J. C. 2013,  arXiv:1310.7012v1\\
Ogilvie, G. I. 1997, MNRAS, 288, 63\\
Ogilvie, G. I., \& Livio, M. 1998, ApJ, 499, 329\\
Ogilvie, G. I, \& Livio, M. 2001, ApJ, 553, 158\\
Ouyed, R., \& Pudritz, R. E. 1999, MNRAS, 309, 233\\
Park, S. J., \& Vishniac, E. T. 1996, ApJ, 471, 158\\
Parker, E. N. 1963, ApJ, 138, 552\\
Penna, R.F., Sadowski, A., Kulkarni, A. K., \& Narayan, R. 2012, MNRAS, 428, 2255\\
Pessah, M. E., Chan, C-K, \& Psaltis, D. 2007, ApJ, 668, L51\\
Pfrommer, C., \& Dursi, L. J. 2010, Nature Physics, 6, 520\\
Pringle, J. E., 1996, Mon. Not. R. Astron. Soc. 281, 357\\
Pringle, J. E., 1997, Mon. Not. R. Astron. Soc. 292, 136\\
Pudritz, R. E, Ouyed, R., Fendt, C., \& Brandenburg, A. 2006, astro-ph/0603592v1\\
Romanova, M. M., Ustyugova, G. V., Koldoba, A. V., Chechetkin, V. M.,
\& Lovelace, R. V. E. 1997, ApJ, 482, 708\\
Rothstein, D. M., \& Lovelace, R. V. E. 2008, ApJ, 677, 1221\\
Rudiger, G. \& Hollerbach, R. 2004, The Magnetic Universe, Geophysical and Astrophysical Dynamo Theory, Wiley-VCH Verlag GmbH \& Co. KGaA\\
Sano, T., Inutsuka, S. I., Turner, N. J., \& Stone, J. M. 2004, ApJ, 605, 321\\
Shakura, N. I., \& Syunyaev, R. A. 1973, A\&A, 24, 337\\
Shi, J., Krolik, J. H., \& Hirose, S. 2010, ApJ, 708, 1716\\
Shore, S. N., LaRosa, T. N. 1999, ApJ, 521, 587\\
Simon, J. B., Beckwith, K., \& Armitage, P. J. 2012, MNRAS, 422, 2685\\
Spitzer, L. 1957, ApJ, 125, 525\\
Spruit, H. C. 1974, Solar Phys. 34, 277\\
Spruit, H. C., \& Uzdensky, D. A. 2005, ApJ, 629, 960\\
Stepiniski, T. F. 1995, Rev. Mix. Astron. Astrofis., 1, 267\\
Suzuki, T. K., \& Inutsuka, S.-i. 2009, ApJL, 691, L49\\
Suzuki, T. K., Muto, T., \& Inutsuka, S.-i. 2010, ApJ, 718, 1289\\
Suzuki, T. K., \& Inutsuka, S. 2014, ApJ, 784, 121\\
Tao, L., Proctor, M. R. E., \& Weiss, N. O. 1998, Mon. Not. R. Astron. Soc. 300, 907\\
Thelen, J-C., Cattaneo, F. 2001, Dynamo and Dynamics, a Mathematical Challenge
NATO Science Series Volume 26, 101\\
Tout, C. A., \& Pringle,  J. E. 1996, MNRAS, 281, 219\\
Ustyugova, G. V., Koldoba, A. V., Romanova, M. M., Chechetkin, V. M., \&
Lovelace, R. V. E. 1999, ApJ, 516, 221\\
Uzdensky, D. A., \& Goodman, J. 2008, ApJ, 682, 608\\
van Ballegooijen A. A., 1989, in Belvedere G., ed., Accretion Disks and
Magnetic Fields in Astrophysics. Kluwer, Dordrecht, p. 99\\
Vainshtein S. I. 1978, Magn. Gidrodin. No. 2, 67\\
Vishniac, E. T., \& Diamond, P. 1992, ApJ, 398, 561\\
Wang, Y.-M. 1987, A\&A, 183, 257\\
Zeldovich Ya. B. 1957, Sov. Phys. JETP 4, 460\\
Ziegler, U., \& R{\"u}diger, G. 2000, A\&A, 356, 1141\\

\end{document}